\documentclass[12pt,a4paper]{article}

\usepackage{authblk}
\usepackage{amsmath,amssymb}
\usepackage{graphicx}
\usepackage{amsfonts}
\usepackage{epsfig}

\usepackage{tikz}
\usepackage{pgflibrarysnakes}
\usetikzlibrary[snakes]
\usetikzlibrary{arrows,shapes,positioning}
\usetikzlibrary{decorations.markings}
\tikzstyle arrowstyle=[scale=1]
\tikzstyle directed=[postaction={decorate,decoration={markings,
		mark=at position .65 with {\arrow[arrowstyle]{stealth}}}}]
\tikzstyle reverse directed=[postaction={decorate,decoration={markings,
		mark=at position .65 with {\arrowreversed[arrowstyle]{stealth};}}}]

\renewcommand{\baselinestretch}{1.5}

\newcommand{\im}{\mathrm i}

\newcommand{\tr}{\operatorname{Tr}}

\newcommand{\ket}[1]{\left|#1\right\rangle}      % Ket-Zustand
\newcommand{\bra}[1]{\left\langle #1\right|}     % Bra-Zustand
\newcommand{\eq}{\begin{equation}}
	\newcommand{\en}{\end{equation}}
\newcommand{\bear}{\begin{eqnarray}}
	\newcommand{\ear}{\end{eqnarray}}

\def\weight[#1][#2][#3][#4][#5]{W\!\!
	\left(\!\!\!\left.\begin{array}{cc}
		#4 \!&\! #3\\#1 \!&\! #2\end{array}\!\right|\!\,#5
	\!\right)}

\def\weightT[#1][#2][#3][#4][#5]{\widetilde{W}
	\left(\left.\begin{array}{cc}
		#4&#3\\#1&#2\end{array}\right|\,#5
	\right)}

\title{\mbox{}Correlation functions of the six-vertex IRF model and its quantum spin chain}
\author{T.S. Tavares
	\footnote{E-mail: tstavares@usp.br}
	\  \ and \ \ G.A.P. Ribeiro\footnote{E-mail: pavan@df.ufscar.br}}
\affil{$^{*}$ Instituto de F\'isica, Universidade de S\~ao Paulo, C.P. 66318, S\~ao Paulo, SP 05315-970, Brazil
\\
$^{\dagger}$ Departamento de F\'{i}sica, Universidade Federal de S\~ao Carlos \\ S\~ao Carlos, SP 13565-905, Brazil\vspace{8pt}}

\date{}

\begin{document}
\renewcommand{\baselinestretch}{1.2}
\maketitle
	
\vspace{-.5in}\noindent
\begin{abstract}
We consider the interaction-round-a-face version of the isotropic six-vertex model. The associated spin chain is made of two coupled Heisenberg spin chains with different boundary twists. The phase diagram of the model and the long distance correlations were studied in [Nucl. Phys. B, 995 (2023) 116333]. Here, we compute the short-distance correlation functions of the model in the ground state for finite system sizes  via non-linear integral equations and in the thermodynamic limit. This was possible since the model satisfies the face version of the discrete quantum Knizhnik-Zamolodchikov (qKZ) equation. A suitable ansatz for the density matrix is proposed in the form of a direct sum of two Heisenberg density matrices, which allows us to obtain the discrete functional equation for the two-site function $\omega(\lambda_1,\lambda_2)$. Thanks to the known results on the factorization of correlation functions of the Heisenberg chain, we are able to compute the density matrix of the IRF model for up to four sites and its associated spin chain for up to three sites.
\end{abstract}

\thispagestyle{empty}

\newpage

\pagestyle{plain}

\pagenumbering{arabic}
\newpage

\renewcommand{\baselinestretch}{1.5}

\section{Introduction}

The correlation functions of integrable models have been widely studied in the last decades \cite{BOOK-KBI,BOOK-JM}. Many results were obtained for quantum spin chains associated to classical vertex models, specially for the $SU(2)$ spin-$1/2$ chain \cite{TAKA,JMMN92,JM96,KMT00,BOKO01,GAS05,BJMST,BST,BGKS06,CORRXXZ,KKMST09,BK2012,MS2018}, its higher-spin
realizations \cite{BoWe94,Idzumi94,Kitanine01,DeMa10,GSS10,KNS2013,RK2016,BS2019} and also some explicit results for high-rank spin chains \cite{BOOS2018,RK2019,RIBEIRO2020}.

Nevertheless, much less is known about correlation functions of the interaction-round-a-face (IRF) models and its associated spin chains \cite{FODA,LUKYANOV,VARCHENKO,NICOLLI2013,TERRAS}. The IRF model and its many different realizations as the cyclic solid-on-solid (CSOS) \cite{BAXTER1973,BAXTER,BAXTER1984,KUNIBA,PEARCE1988}, the restricted solid-on-solid (RSOS) models and its $A$-$D$-$E$ generalizations \cite{PEARCE1989,PASQUIER,PEARCE1992,PEARCE1993} share some similarities with the vertex models and its integrable structure, which allow for the exact computation of physical properties. Recently, some remarkable results appeared in the context of correlation functions of face models \cite{FRAHM2021,FRAHM2023}. In \cite{FRAHM2021}, the reduced density matrix was formulated in terms of face model weights, which allowed for the derivation of discrete functional equations of qKZ type along the same lines as the six-vertex model and the associated Heisenberg spin chain \cite{BK2012}. Besides, it was shown \cite{FRAHM2021,FRAHM2023} that the density matrix of the RSOS models can be factorized in terms of nearest-neighbour correlators.

In this work, we are interested to compute correlation functions of the recently proposed interaction-round-a-face version of the six-vertex model \cite{TAVARES2023}. The spin chain associated to this face model have previously appeared in different contexts \cite{POZSGAY,H6v,FENDLEY}, however, the integrable structure of the face models and its relationship with the six-vertex model allowed the study of its physical properties and phase diagram in the thermodynamical limit \cite{TAVARES2023}. Here, inspired in \cite{FRAHM2021,FRAHM2023}, we study the reduced density matrix for the interaction-round-a-face version of the six-vertex model, which satisfies the discrete functional equation of qKZ type. We obtain the solution of the functional equation at zero temperature for finite and infinity system sizes, which is determined by the density matrix in the ground state.
As described in \cite{TAVARES2023}, the associated quantum spin chain is made of two coupled Heisenberg spin chains with different boundary conditions. This implies that the density matrix in the ground state can also be seen as two copies of the density matrix of the Heisenberg spin chain. Therefore, with a suitable ansatz, we obtain that the qKZ equation for the face model density matrix results in the discrete functional equation for the two-site correlation function, usually denote by $\omega(\lambda_1,\lambda_2)$ and whose solution is written in terms of the solution of non-linear integral equations for finite system sizes. This allowed to fully determine the two and three-site density matrices of the spin chain and its correlations at zero temperature for finite system size and in the thermodynamic limit.

This paper is organized as follows. In section \ref{integrability}, we describe the IRF version of the six-vertex model and its integrable structure.  In section \ref{secdensity}, we introduce the physical density matrix and its functional equation. In section \ref{res}, we present the solution for two-, three-, and four-site density matrices of the IRF  model and up to three-sites for the spin chain for finite system size and in the thermodynamic limit at zero temperature. In section \ref{NLIEsec}, we make  use of the non-linear integral equations in order to evaluate the non-trivial correlations for large but finite system sizes. Our conclusions are given in section \ref{CONCLUSION}.

\section{The IRF version of the six-vertex model and its quantum spin chain}\label{integrability}

The face models are classical statistical mechanical models on a square lattice defined by local Boltzmann weights, which can be depicted as \cite{BAXTER1973,BAXTER,BAXTER1984},
\eq
\begin{aligned}
	\begin{tikzpicture}[scale=1.25]
		\draw (0.0,0.) node {$\weight[a][b][c][d][\lambda]=$};
		\draw (2.0,0) [-,color=black, thick]	+(-0.5,-0.5) -- +(-0.5,0.5)-- +(0.5,0.5) -- +(0.5,-0.5)-- +(-0.5,-0.5);
		\draw (1.45,-0.66) node {$a$};
		\draw (2.55,-0.66) node {$b$};
		\draw (2.55, 0.64) node {$c$};
		\draw (1.45, 0.65) node {$d$};
		\draw (2., 0.) node {$\lambda$};
		\draw (2.,0) [-,color=black,  thick, rounded corners=7pt]	+(-0.3,-0.5) -- +(-0.3,-0.3) -- +(-0.5,-0.3) ;
		\draw (2.75, 0.) node {,};
	\end{tikzpicture}
\end{aligned}
\en
where $a,b,c,d$ are the spins or heights of the corners of the face separated by bonds and $\lambda$ is the spectral parameter.

\begin{figure}[thb]
	\begin{minipage}{\linewidth}
		\begin{center}
			\begin{tikzpicture}[scale=1.55]
				\draw (0,1.5) [-,color=black, dotted, directed, rounded corners=7pt]+(-(0.,0) --+(0.,0.5);
				\draw (0,1.5) [-,color=black, dotted, directed, rounded corners=7pt]+(-(0.,-0.5) --+(0.,0.);
				\draw (0,1.5) [-,color=black,  dotted, directed, rounded corners=7pt]	+(0,0) -- +(0.5,0) ;
				\draw (0,1.5) [-,color=black,  dotted, directed, rounded corners=7pt]	+(-0.5,0) -- +(0.,0) ;
				\draw (0,1.5) [-,color=black,  thick, rounded corners=7pt]	+(-0.3,-0.5) -- +(-0.3,-0.3) -- +(-0.5,-0.3) ;
				\draw (0,1.5) [-,color=black, thick]	+(-0.5,-0.5) -- +(-0.5,0.5)-- +(0.5,0.5) -- +(0.5,-0.5)--+(-0.5,-0.5);
				\draw (-0.6,-0.6+1.5) node {\tiny $+$};
				\draw (0.6,-0.6+1.5) node {\tiny $+$};
				\draw (0.6, 0.6+1.5) node {\tiny$+$};
				\draw (-0.6, 0.6+1.5) node {\tiny$+$};
				\draw (0.75, 0.9+1.75) node {$\frak{a}(\lambda)$};
				\draw (2.25,0) [-,color=black,  thick, rounded corners=7pt]	+(0,-1) -- +(0,3);

				\draw (1.5,1.5) [-,color=black, dotted, directed, rounded corners=7pt]+(-(0.,0.5) --+(0.,0);
				\draw (1.5,1.5) [-,color=black, dotted, directed, rounded corners=7pt]+(-(0.,0) --+(0.,-0.5);
				\draw (1.5,1.5) [-,color=black,  dotted, directed, rounded corners=7pt]	+(0.5,0) -- +(0,0) ;
				\draw (1.5,1.5) [-,color=black,  dotted, directed, rounded corners=7pt]	+(0,0) -- +(-0.5,0) ;
				\draw (1.5,1.5) [-,color=black,  thick, rounded corners=7pt]	+(-0.3,-0.5) -- +(-0.3,-0.3) -- +(-0.5,-0.3) ;
				\draw (1.5,1.5) [-,color=black, thick]	+(-0.5,-0.5) -- +(-0.5,0.5)-- +(0.5,0.5) -- +(0.5,-0.5)--+(-0.5,-0.5);
				\draw (0.95,-0.6+1.5) node {\tiny$-$};
				\draw (2.1,-0.6+1.5) node {\tiny$+$};
				\draw (2.1, 0.6+1.5) node {\tiny$-$};
				\draw (0.95, 0.6+1.5) node {\tiny$+$};
				
				\draw (0,0) [-,color=black, dotted, directed, rounded corners=7pt]+(-(0.,0) --+(0.,0.5);
				\draw (0,0) [-,color=black, dotted, directed, rounded corners=7pt]+(-(0.,-0.5) --+(0.,0);
				\draw (0,0) [-,color=black,  dotted, directed, rounded corners=7pt]	+(0,0) -- +(0.5,0) ;
				\draw (0,0) [-,color=black,  dotted, directed, rounded corners=7pt]	+(-0.5,0) -- +(0,0) ;
				\draw (0,0) [-,color=black,  thick, rounded corners=7pt]	+(-0.3,-0.5) -- +(-0.3,-0.3) -- +(-0.5,-0.3) ;
				\draw (0,0) [-,color=black, thick]	+(-0.5,-0.5) -- +(-0.5,0.5)-- +(0.5,0.5) -- +(0.5,-0.5)--+(-0.5,-0.5);
				\draw (-0.55,-0.6) node {\tiny $-$};
				\draw (0.6,-0.6) node {\tiny $-$};
				\draw (0.6, 0.6) node {\tiny $-$};
				\draw (-0.55, 0.6) node {\tiny $-$};
				
				\draw (1.5,0) [-,color=black, dotted, directed, rounded corners=7pt]+(-(0.,0.5) --+(0.,0);
				\draw (1.5,0) [-,color=black, dotted, directed, rounded corners=7pt]+(-(0.,0) --+(0.,-0.5);
				\draw (1.5,0) [-,color=black,  dotted, directed, rounded corners=7pt]	+(0.5,0) -- +(0,0) ;
				\draw (1.5,0) [-,color=black,  dotted, directed, rounded corners=7pt]	+(0,0) -- +(-0.5,0) ;
				\draw (1.5,0) [-,color=black,  thick, rounded corners=7pt]	+(-0.3,-0.5) -- +(-0.3,-0.3) -- +(-0.5,-0.3) ;
				\draw (1.5,0) [-,color=black, thick]	+(-0.5,-0.5) -- +(-0.5,0.5)-- +(0.5,0.5) -- +(0.5,-0.5)--+(-0.5,-0.5);
				\draw (0.95,-0.6) node {\tiny$+$};
				\draw (2.1,-0.6) node {\tiny$-$};
				\draw (2.1, 0.6) node {\tiny$+$};
				\draw (0.95, 0.6) node {\tiny$-$};
				
				%%%%%%%%%%%%%%%%%%%%%%%%%%%
				
				\draw (3,1.5) [-,color=black, dotted, directed, rounded corners=7pt]+(-(0.,0) --+(0.,0.5);
				\draw (3,1.5) [-,color=black, dotted, directed, rounded corners=7pt]+(-(0.,-0.5) --+(0.,0.);
				\draw (3,1.5) [-,color=black,  dotted, directed, rounded corners=7pt]	+(0.5,0) -- +(0,0) ;
				\draw (3,1.5) [-,color=black,  dotted, directed, rounded corners=7pt]	+(0,0) -- +(-0.5,0) ;
				\draw (3,1.5) [-,color=black,  thick, rounded corners=7pt]	+(-0.3,-0.5) -- +(-0.3,-0.3) -- +(-0.5,-0.3) ;
				\draw (3,1.5) [-,color=black, thick]	+(-0.5,-0.5) -- +(-0.5,0.5)-- +(0.5,0.5) -- +(0.5,-0.5)--+(-0.5,-0.5);
				\draw (-0.6+3,-0.6+1.5) node {\tiny $+$};
				\draw (0.6+3,-0.6+1.5) node {\tiny $+$};
				\draw (0.6+3, 0.6+1.5) node {\tiny$-$};
				\draw (-0.6+3, 0.6+1.5) node {\tiny$-$};

				\draw (3.75, 0.9+1.75) node {$\frak{b}(\lambda)$};
				\draw (5.25,0) [-,color=black,  thick, rounded corners=7pt]	+(0,-1) -- +(0,3);

				\draw (4.5,1.5) [-,color=black, dotted, directed, rounded corners=7pt]+(-(0.,0.5) --+(0.,0);
				\draw (4.5,1.5) [-,color=black, dotted, directed, rounded corners=7pt]+(-(0.,0) --+(0.,-0.5);
				\draw (4.5,1.5) [-,color=black,  dotted, directed, rounded corners=7pt]	+(0,0) -- +(0.5,0) ;
				\draw (4.5,1.5) [-,color=black,  dotted, directed, rounded corners=7pt]	+(-0.5,0) -- +(0,0) ;
				\draw (4.5,1.5) [-,color=black,  thick, rounded corners=7pt]	+(-0.3,-0.5) -- +(-0.3,-0.3) -- +(-0.5,-0.3) ;
				\draw (4.5,1.5) [-,color=black, thick]	+(-0.5,-0.5) -- +(-0.5,0.5)-- +(0.5,0.5) -- +(0.5,-0.5)--+(-0.5,-0.5);
				\draw (0.95+3,-0.6+1.5) node {\tiny$+$};
				\draw (5.15,-0.6+1.5) node {\tiny$-$};
				\draw (5.15, 0.6+1.5) node {\tiny$-$};
				\draw (3.95, 0.6+1.5) node {\tiny$+$};
				
				\draw (3,0) [-,color=black, dotted, directed, rounded corners=7pt]+(-(0.,0) --+(0.,0.5);
				\draw (3,0) [-,color=black, dotted, directed, rounded corners=7pt]+(-(0.,-0.5) --+(0.,0);
				\draw (3,0) [-,color=black,  dotted, directed, rounded corners=7pt]	+(0.5,0) -- +(0,0) ;
				\draw (3,0) [-,color=black,  dotted, directed, rounded corners=7pt]	+(0,0) -- +(-0.5,0);
				\draw (3,0) [-,color=black,  thick, rounded corners=7pt]	+(-0.3,-0.5) -- +(-0.3,-0.3) -- +(-0.5,-0.3) ;
				\draw (3,0) [-,color=black, thick]	+(-0.5,-0.5) -- +(-0.5,0.5)-- +(0.5,0.5) -- +(0.5,-0.5)--+(-0.5,-0.5);
				\draw (-0.55+3,-0.6) node {\tiny $-$};
				\draw (3.6,-0.6) node {\tiny $-$};
				\draw (3.6, 0.6) node {\tiny $+$};
				\draw (-0.55+3, 0.6) node {\tiny $+$};
				
				\draw (4.5,0) [-,color=black, dotted, directed, rounded corners=7pt]+(-(0.,0.5) --+(0.,0);
				\draw (4.5,0) [-,color=black, dotted, directed, rounded corners=7pt]+(-(0.,0) --+(0.,-0.5);
				\draw (4.5,0) [-,color=black,  dotted, directed, rounded corners=7pt]	+(0,0) -- +(0.5,0) ;
				\draw (4.5,0) [-,color=black,  dotted, directed, rounded corners=7pt]	+(-0.5,0) -- +(0,0) ;
				\draw (4.5,0) [-,color=black,  thick, rounded corners=7pt]	+(-0.3,-0.5) -- +(-0.3,-0.3) -- +(-0.5,-0.3) ;
				\draw (4.5,0) [-,color=black, thick]	+(-0.5,-0.5) -- +(-0.5,0.5)-- +(0.5,0.5) -- +(0.5,-0.5)--+(-0.5,-0.5);
				\draw (3.95,-0.6) node {\tiny$-$};
				\draw (5.15,-0.6) node {\tiny$+$};
				\draw (5.15, 0.6) node {\tiny$+$};
				\draw (3.95, 0.6) node {\tiny$-$};
		
				%%%%%%%%%%%%%%%%%%%%%%%%%%%
				
				\draw (6,1.5) [-,color=black, dotted, directed, rounded corners=7pt]+(-(0.,0.5) --+(0.,0);
				\draw (6,1.5) [-,color=black, dotted, directed, rounded corners=7pt]+(-(0.,-0.5) --+(0.,0.);
				\draw (6,1.5) [-,color=black,  dotted, directed, rounded corners=7pt]	+(0,0) -- +(0.5,0) ;
				\draw (6,1.5) [-,color=black,  dotted, directed, rounded corners=7pt]	+(0,0) -- +(-0.5,0) ;
				\draw (6,1.5) [-,color=black,  thick, rounded corners=7pt]	+(-0.3,-0.5) -- +(-0.3,-0.3) -- +(-0.5,-0.3) ;
				\draw (6,1.5) [-,color=black, thick]	+(-0.5,-0.5) -- +(-0.5,0.5)-- +(0.5,0.5) -- +(0.5,-0.5)--+(-0.5,-0.5);
				\draw (-0.6+6,-0.6+1.5) node {\tiny $+$};
				\draw (0.6+6,-0.6+1.5) node {\tiny $+$};
				\draw (0.6+6, 0.6+1.5) node {\tiny$+$};
				\draw (-0.6+6, 0.6+1.5) node {\tiny$-$};

				\draw (6.75, 0.9+1.75) node {$\frak{c}(\lambda)$};

				\draw (7.5,1.5) [-,color=black, dotted, directed, rounded corners=7pt]+(-(0.,0) --+(0.,0.5);
				\draw (7.5,1.5) [-,color=black, dotted, directed, rounded corners=7pt]+(-(0.,0) --+(0.,-0.5);
				\draw (7.5,1.5) [-,color=black,  dotted, directed, rounded corners=7pt]	+(0.5,0) -- +(0,0) ;
				\draw (7.5,1.5) [-,color=black,  dotted, directed, rounded corners=7pt]	+(-0.5,0) -- +(0,0) ;
				\draw (7.5,1.5) [-,color=black,  thick, rounded corners=7pt]	+(-0.3,-0.5) -- +(-0.3,-0.3) -- +(-0.5,-0.3) ;
				\draw (7.5,1.5) [-,color=black, thick]	+(-0.5,-0.5) -- +(-0.5,0.5)-- +(0.5,0.5) -- +(0.5,-0.5)--+(-0.5,-0.5);
				\draw (0.95+6,-0.6+1.5) node {\tiny$+$};
				\draw (8.15,-0.6+1.5) node {\tiny$-$};
				\draw (8.15, 0.6+1.5) node {\tiny$+$};
				\draw (6.95, 0.6+1.5) node {\tiny$+$};
				
				\draw (6,0) [-,color=black, dotted, directed, rounded corners=7pt]+(-(0.,0.5) --+(0.,0);
				\draw (6,0) [-,color=black, dotted, directed, rounded corners=7pt]+(-(0.,-0.5) --+(0.,0);
				\draw (6,0) [-,color=black,  dotted, directed, rounded corners=7pt]	+(0,0) -- +(0.5,0) ;
				\draw (6,0) [-,color=black,  dotted, directed, rounded corners=7pt]	+(0,0) -- +(-0.5,0);
				\draw (6,0) [-,color=black,  thick, rounded corners=7pt]	+(-0.3,-0.5) -- +(-0.3,-0.3) -- +(-0.5,-0.3) ;
				\draw (6,0) [-,color=black, thick]	+(-0.5,-0.5) -- +(-0.5,0.5)-- +(0.5,0.5) -- +(0.5,-0.5)--+(-0.5,-0.5);
				\draw (-0.55+6,-0.6) node {\tiny $-$};
				\draw (6.6,-0.6) node {\tiny $-$};
				\draw (6.6, 0.6) node {\tiny $-$};
				\draw (-0.55+6, 0.6) node {\tiny $+$};
				
				\draw (7.5,0) [-,color=black, dotted, directed, rounded corners=7pt]+(-(0.,0) --+(0.,0.5);
				\draw (7.5,0) [-,color=black, dotted, directed, rounded corners=7pt]+(-(0.,0) --+(0.,-0.5);
				\draw (7.5,0) [-,color=black,  dotted, directed, rounded corners=7pt]	+(0.5,0) -- +(0,0) ;
				\draw (7.5,0) [-,color=black,  dotted, directed, rounded corners=7pt]	+(-0.5,0) -- +(0,0) ;
				\draw (7.5,0) [-,color=black,  thick, rounded corners=7pt]	+(-0.3,-0.5) -- +(-0.3,-0.3) -- +(-0.5,-0.3) ;
				\draw (7.5,0) [-,color=black, thick]	+(-0.5,-0.5) -- +(-0.5,0.5)-- +(0.5,0.5) -- +(0.5,-0.5)--+(-0.5,-0.5);
				\draw (6.95,-0.6) node {\tiny$-$};
				\draw (8.15,-0.6) node {\tiny$+$};
				\draw (8.15, 0.6) node {\tiny$-$};
				\draw (6.95, 0.6) node {\tiny$-$};
				
			\end{tikzpicture}
		\end{center}
	\end{minipage}
	\caption{Boltzmann weights of the IRF version of the six-vertex model. The face weights are obtained from the allowed configuration for the six-vertex model, which can be seen through the dotted oriented lines.}
	\label{IRF6V}
\end{figure}
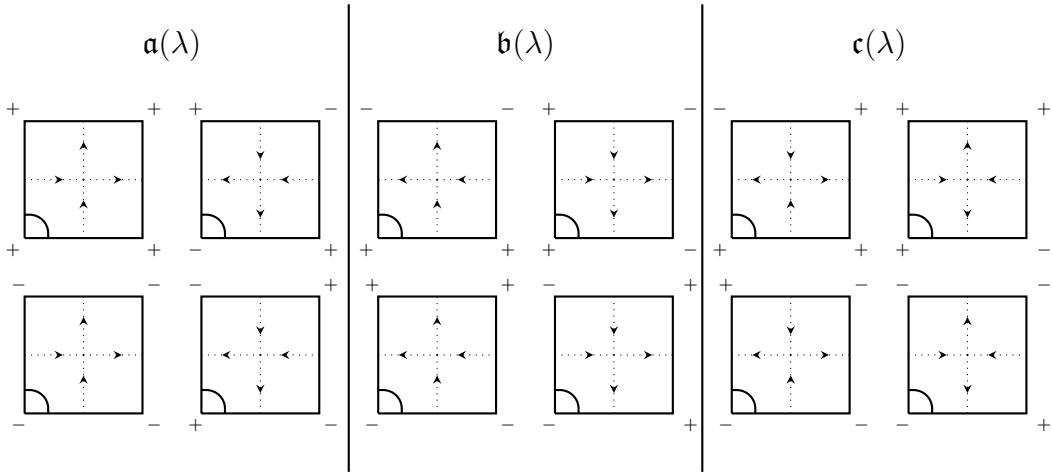

In this work, we consider the IRF version of the isotropic six-vertex model introduced in  \cite{TAVARES2023}, which can be depicted as in Figure \ref{IRF6V}. This face model is made of two copies of the six-vertex model, whose spins are assigned as in Figure \ref{IRF6V}.

The face weights $\mathfrak{a}(\lambda)$, $\mathfrak{b}(\lambda)$ and  $\mathfrak{c}(\lambda)$ are given by,
\begin{alignat}{5}
	\weight[+][+][+][+][\lambda] \!&=	\weight[-][-][-][-][\lambda] \!&=	\weight[-][+][-][+][\lambda] \!&=	\weight[+][-][+][-][\lambda] &=\mathfrak{a}(\lambda)&=\lambda+1, \nonumber \\
	\weight[+][+][-][-][\lambda] \!&=\weight[-][-][+][+][\lambda] \!&=\weight[+][-][-][+][\lambda] \!&=\weight[-][+][+][-][\lambda] &=\mathfrak{b}(\lambda)&=\lambda, \label{wIRF6V} \\
	\weight[+][+][+][-][\lambda] \!&=\weight[-][-][-][+][\lambda] \!&=\weight[+][-][+][+][\lambda] \!&=\weight[-][+][-][-][\lambda] &=\mathfrak{c}(\lambda)&=1. \nonumber
\end{alignat}

The physical properties of the classical $M\times L$ square lattice model with periodic boundary condition
can be obtained from the partition function, which can be written as $Z_{\text{IRF}}=\tr{\left[\left(T_{\text{IRF}}(\lambda)\right)^M\right]}$. Here, the transfer matrix with periodic boundary condition is defined as $T_{\text{IRF}}(\lambda)=\tr[{\cal T}(\lambda)]=\sum_{\alpha_1,\alpha_2} {\cal T}^{\alpha_1 \alpha_1}_{\alpha_2\alpha_2}(\lambda)$, where the monodromy matrix elements are defined as the product of the Boltzmann weights along the row,
\bear
{{\cal T}^{\alpha_1\beta_1}_{\alpha_2 \beta_2}}(\lambda)^{a_1 a_2 \cdots a_L}_{b_1 b_2 \cdots b_L}=\prod_{i=1}^L \weight[b_{i}][b_{i+1}][a_{i+1}][a_{i}][\lambda-u_i]\delta_{\alpha_1 a_1}\delta_{\alpha_2 b_1}\delta_{\beta_1 a_{L+1}}\delta_{\beta_2 b_{L+1}},
\label{monodromy}
\ear
where for later convenience we introduce the inhomogeneity parameters $\{u_i\}$.

The depiction of the transfer matrix is given in Figure \ref{monodromyFig}.
\begin{figure}[h]
	\begin{minipage}{\linewidth}
		\begin{center}
			\begin{tikzpicture}[scale=1.5]
				\draw (-0.650,0.) node {${\cal T}^{\alpha_1\beta_1}_{\alpha_2 \beta_2}(\lambda)^{a_1 a_2 \cdots a_L}_{b_1 b_2 \cdots b_L}=$};
				\draw (1.0,0) [-,color=black, thick]	+(-0.5,-0.5) -- +(-0.5,0.5)-- +(0.5,0.5) -- +(0.5,-0.5)-- +(-0.5,-0.5);
				\draw (2.0,0) [-,color=black, thick]	+(-0.5,-0.5) -- +(-0.5,0.5)-- +(0.5,0.5) -- +(0.5,-0.5)-- +(-0.5,-0.5);
				\draw (3.0,0) [-,color=black, thick]	+(-0.5,-0.5) -- +(-0.5,0.5)-- +(0.5,0.5) -- +(0.5,-0.5)-- +(-0.5,-0.5);
				\draw (4.0,0) [-,color=black, thick]	+(-0.5,-0.5) -- +(-0.5,0.5)-- +(0.5,0.5) -- +(0.5,-0.5)-- +(-0.5,-0.5);
				\draw (5.0,0) [-,color=black, thick]	+(-0.5,-0.5) -- +(-0.5,0.5)-- +(0.5,0.5) -- +(0.5,-0.5)-- +(-0.5,-0.5);
				
				\draw (6.0,0) [-,color=black, thick]	+(-0.5,-0.5) -- +(-0.5,0.5)-- +(0.5,0.5) -- +(0.5,-0.5)-- +(-0.5,-0.5);
				
				\draw (1.01,0.01) [-,color=black,  rounded corners=7pt,thick]	+(-0.35,-0.5) -- +(-0.35,-0.35)-- +(-0.5,-0.35);
				\draw (2.01,0.01) [-,color=black,  rounded corners=7pt,thick]	+(-0.35,-0.5) -- +(-0.35,-0.35)-- +(-0.5,-0.35);
				\draw (3.01,0.01) [-,color=black,  rounded corners=7pt,thick]	+(-0.35,-0.5) -- +(-0.35,-0.35)-- +(-0.5,-0.35);
				\draw (4.01,0.01) [-,color=black,  rounded corners=7pt,thick]	+(-0.35,-0.5) -- +(-0.35,-0.35)-- +(-0.5,-0.35);
				\draw (6.01,0.01) [-,color=black,  rounded corners=7pt,thick]	+(-0.35,-0.5) -- +(-0.35,-0.35)-- +(-0.5,-0.35);
				
				\draw (0.45,-0.65) node {$\alpha_2=b_1$};
				\draw (1.55,-0.65) node {$b_2$};
				\draw (2.55,-0.65) node {$b_3$};
				\draw (3.55,-0.65) node {$b_4$};
				\draw (4.55,-0.65) node {$b_5$};
				\draw (5.65,-0.65) node {$b_{L}$};
				\draw (6.75,-0.66) node {$b_{L+1}=\beta_2$};
				\draw (0.45, 0.75) node {$\alpha_1=a_1$};
				\draw (1.55,0.75) node {$a_2$};
				\draw (2.55,0.75) node {$a_3$};
				\draw (3.55,0.75) node {$a_4$};
				\draw (4.55,0.75) node {$a_5$};
				\draw (5.65,0.75) node {$a_{L}$};
				\draw (6.75, 0.75) node {$a_{L+1}=\beta_1$};
				\draw (1., 0.) node {$\lambda-u_1$};
				\draw (2., 0.) node {$\lambda-u_2$};
				\draw (3., 0.) node {$\lambda-u_3$};
				\draw (4., 0.) node {$\lambda-u_4$};
				\draw (6., 0.) node {$\lambda-u_L$};
				\draw (5, 0.) node {$\cdots$};
			\end{tikzpicture}
		\end{center}
	\end{minipage}
	\caption{The monodromy matrix elements of the IRF model.}
	\label{monodromyFig}
\end{figure}

The transfer matrix is part of a family of commuting operators $[T_{\text{IRF}}(\lambda),T_{\text{IRF}}(\mu)]=0$ thanks to the Yang-Baxter equation. The face version of the Yang-Baxter equation given by,
\bear
\sum_{i} \weight[a][b][i][f][\lambda-\mu]\weight[i][d][e][f][\mu]\weight[b][c][d][i][\lambda]\nonumber\\
=\sum_{i} \weight[a][i][e][f][\lambda]\weight[b][c][i][a][\mu]\weight[i][c][d][e][\lambda-\mu].
\ear

Taking the logarithmic derivative of the IRF transfer matrix ${\cal H}_{\text{IRF}}=\partial_{\lambda}\log T_{\text{IRF}}(\lambda) |_{\lambda=u_i=0}$, we obtain a one-dimensional spin chain with interaction of three spins,
\bear
{\cal H}_{\text{IRF}}= \frac{1}{2}
\sum_{i=1}^L  \sigma_i^x -\sigma_{i-1}^z\sigma_i^x \sigma_{i+1}^z + \sigma_{i-1}^z \sigma_{i+1}^z+1,
\label{Hirf}
\ear
where $\sigma^{\alpha}$ for $\alpha=x,y,z$ are the standard Pauli matrices. The Hamiltonian (\ref{Hirf})  has a continuous $U(1)$ symmetry and a discrete $\mathbb{Z}_2$ symmetry, since it commutes with the operators,
\eq
\Sigma^z= \sum_{j=1}^L \sigma_j^z \sigma_{j+1}^z,\qquad \Pi^x=\prod_{j=1}^L \sigma_j^x. \label{symmetries}
\en

This three spins interaction Hamiltonian (\ref{Hirf}) was shown in \cite{TAVARES2023} to be made of two coupled Heisenberg spin chains (${\cal H}_{\text{XXX}}(\phi)$) with different boundary conditions at the sector of even spin flips,
\eq
{\cal H}_{\text{IRF}} = U^{t} \left( {\cal H}_{\text{XXX}}^{\text{even}}(0) \oplus {\cal H}_{\text{XXX}}^{\text{even}}(\pi/2) \right)U,
\en
where $U$ is the matrix that diagonalizes $\Pi^x$ and $\phi$ is the twist angle, such that $\phi=0$ result in periodic boundary condition \cite{TAVARES2023}.

The physical properties of this model were studied in \cite{TAVARES2023} via the quantum transfer matrix approach\cite{KLUMPER92} in the thermodynamical limit. It was shown that the leading eigenvalue of the transfer matrix with periodic boundary determine thermodynamic properties as free-energy and its derivatives. This implies that the Hamiltonian (\ref{Hirf}) has the same phase diagram as the Heisenberg model. Nevertheless due to the fact that the first excited states belong to the sector with non-periodic boundary condition, the long distance correlation function presents different oscillatory behaviour from the Heisenberg chain \cite{KLUMPER93}.

\section{Density matrix and functional equations}\label{secdensity}

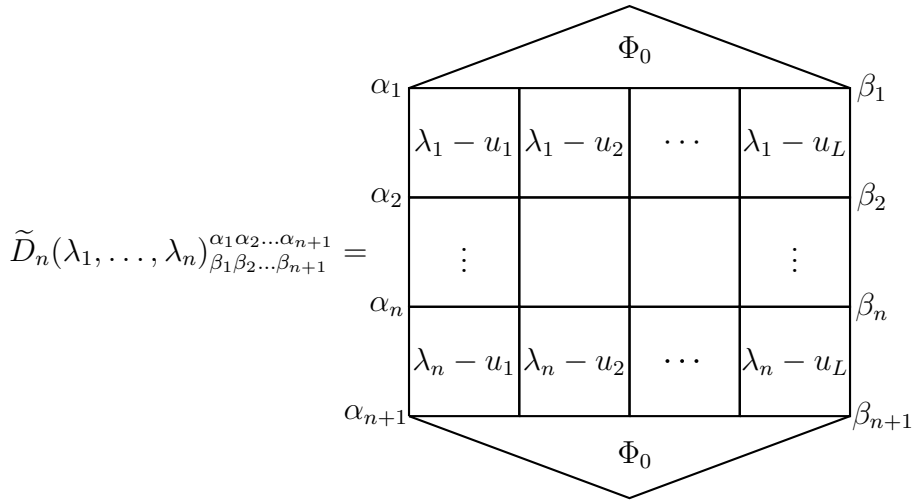
\begin{figure}[h]
	\begin{minipage}{\linewidth}
		\begin{center}
			\begin{tikzpicture}[scale=1.45]
				\draw (-1.50,1.) node {$\widetilde{D}_n(\lambda_1,\dots,\lambda_n)^{\alpha_1\alpha_2\dots\alpha_{n+1}}_{\beta_1\beta_2\dots\beta_{n+1}}=$};
				\draw (1.0,2) [-,color=black, thick]	+(-0.5,0.5) -- +(1.5,1.25)-- +(3.5,0.5);
				\draw (1.0,0) [-,color=black, thick]	+(-0.5,-0.5) -- +(1.5,-1.25)-- +(3.5,-0.5);
				
				\draw (1.0,2) [-,color=black, thick]	+(-0.5,-0.5) -- +(-0.5,0.5)-- +(0.5,0.5) -- +(0.5,-0.5)-- +(-0.5,-0.5);
				\draw (2.0,2) [-,color=black, thick]	+(-0.5,-0.5) -- +(-0.5,0.5)-- +(0.5,0.5) -- +(0.5,-0.5)-- +(-0.5,-0.5);
				\draw (3.0,2) [-,color=black, thick]	+(-0.5,-0.5) -- +(-0.5,0.5)-- +(0.5,0.5) -- +(0.5,-0.5)-- +(-0.5,-0.5);
				\draw (4.0,2) [-,color=black, thick]	+(-0.5,-0.5) -- +(-0.5,0.5)-- +(0.5,0.5) -- +(0.5,-0.5)-- +(-0.5,-0.5);

				\draw (1.0,1) [-,color=black, thick]	+(-0.5,-0.5) -- +(-0.5,0.5)-- +(0.5,0.5) -- +(0.5,-0.5)-- +(-0.5,-0.5);
				\draw (2.0,1) [-,color=black, thick]	+(-0.5,-0.5) -- +(-0.5,0.5)-- +(0.5,0.5) -- +(0.5,-0.5)-- +(-0.5,-0.5);
				\draw (3.0,1) [-,color=black, thick]	+(-0.5,-0.5) -- +(-0.5,0.5)-- +(0.5,0.5) -- +(0.5,-0.5)-- +(-0.5,-0.5);
				\draw (4.0,1) [-,color=black, thick]	+(-0.5,-0.5) -- +(-0.5,0.5)-- +(0.5,0.5) -- +(0.5,-0.5)-- +(-0.5,-0.5);

				\draw (1.0,0) [-,color=black, thick]	+(-0.5,-0.5) -- +(-0.5,0.5)-- +(0.5,0.5) -- +(0.5,-0.5)-- +(-0.5,-0.5);
				\draw (2.0,0) [-,color=black, thick]	+(-0.5,-0.5) -- +(-0.5,0.5)-- +(0.5,0.5) -- +(0.5,-0.5)-- +(-0.5,-0.5);
				\draw (3.0,0) [-,color=black, thick]	+(-0.5,-0.5) -- +(-0.5,0.5)-- +(0.5,0.5) -- +(0.5,-0.5)-- +(-0.5,-0.5);
				\draw (4.0,0) [-,color=black, thick]	+(-0.5,-0.5) -- +(-0.5,0.5)-- +(0.5,0.5) -- +(0.5,-0.5)-- +(-0.5,-0.5);
				
				\draw (2.55, 2.85) node {$\Phi_0$};
				\draw (2.55, -0.85) node {$\Phi_0$};				
								
				\draw (0.3, 1.5) node {$\alpha_2$};
				\draw (4.7, 1.5) node {$\beta_2$};
				\draw (0.3, 2.5) node {$\alpha_1$};
				\draw (4.7, 2.5) node {$\beta_1$};

				\draw (0.2,-0.5) node {$\alpha_{n+1}$};
				\draw (4.8,-0.5) node {$\beta_{n+1}$};
				\draw (0.3, 0.5) node {$\alpha_n$};
				\draw (4.7, 0.5) node {$\beta_n$};

				\draw (1., 2.) node {$\lambda_1-u_1$};
				\draw (2., 2.) node {$\lambda_1-u_2$};
				\draw (4., 2.) node {$\lambda_1-u_L$};
				\draw (3, 2.) node {$\cdots$};
				\draw (1., 1.) node {$\vdots$};
				\draw (4., 1.) node {$\vdots$};
				\draw (3, 0.) node {$\cdots$};
				\draw (1., 0.) node {$\lambda_n-u_1$};
				\draw (2., 0.) node {$\lambda_n-u_2$};
				\draw (4., 0.) node {$\lambda_n-u_L$};
				\draw (3, 0.) node {$\cdots$};

			\end{tikzpicture}
		\end{center}
	\end{minipage}
	\caption{Graphical illustration of the un-normalized reduced density matrix elements.}
	\label{Dn}
\end{figure}

In \cite{FRAHM2021}, the scheme to deal with correlation functions of integrable models was extended to the case of face models. This was done by proceeding along the same lines as in the vertex model case \cite{BK2012}. Within this approach, the main object is the inhomogeneous reduced density matrix formulated in terms of face weights at zero temperature and finite system size $L$ (see Figure \ref{Dn}), whose matrix element are given by,
\eq
D_n(\lambda_1,\lambda_2,\dots,\lambda_n)^{\alpha_1\alpha_2\dots\alpha_{n+1}}_{\beta_1\beta_2\dots\beta_{n+1}}=\frac{\bra{\Phi_0} {\cal T}^{\alpha_1\beta_1}_{\alpha_2\beta_2}(\lambda_1) {\cal T}^{\alpha_2\beta_2}_{\alpha_3\beta_3}(\lambda_2)\dots {\cal T}^{\alpha_n\beta_n}_{\alpha_{n+1}\beta_{n+1}}(\lambda_n) \ket{\Phi_0}}{\langle \Phi_0 \ket{\Phi_0}\Lambda_0(\lambda_1)\Lambda_0(\lambda_2)\cdots \Lambda_0(\lambda_n)},
\label{density}
\en
where ${{\cal T}^{\alpha_k\beta_k}_{\alpha_{k+1} \beta_{k+1}}}(\lambda)$ are the monodromy matrix elements (\ref{monodromy}) and $\ket{\Phi_0}$ is the eigenstate associated to the leading eigenvalue  $\Lambda_0(x)$ of the transfer matrix $T_{\text{IRF}}(\lambda)$.

It is worth to note that the connection with the physical density matrix of the face model for $n$ consecutive edges in the ground state is obtained from the inhomogeneous reduced density matrix via the limit $\lambda_k \to u_k$ for $k=1,2,\dots,n$ \cite{FRAHM2021},
\eq
\check{D}_{[1,n]}= \lim_{\lambda_1\to u_1,\cdots,\lambda_n\rightarrow u_n} D_n(\lambda_1,\lambda_2,\cdots,\lambda_n).
\en
Besides, in order to obtain the physical results for the quantum spin chain, one has to additionally take partial trace of the IRF density matrix, which results in the spin chain density matrix ${\mathbb D}_{n-1}$ given by,
\eq
{\mathbb D}_{n-1}(\lambda_1,\lambda_2,\cdots,\lambda_n)^{\alpha_2\dots\alpha_{n}}_{\beta_2\dots\beta_{n}} =\sum_{\alpha_1,\alpha_{n+1}} D_n(\lambda_1,\lambda_2,\cdots,\lambda_n)^{\alpha_1\alpha_2\dots\alpha_n\alpha_{n+1}}_{\alpha_1\beta_2\dots\beta_n\alpha_{n+1}}.
\en
Finally, the homogeneous limit guarantees we are describing the correlation of the quantum spin chain, which means that all inhomogeneities are taken to zero such that $u_k=0$, for all $k$.

The efficient computation of the inhomogeneous reduced density matrix in a way that the above limits can be taken is established through the solution of a discrete functional equation of qKZ type \cite{BK2012,FRAHM2021}. The existence of such equation is guaranteed by the integrable structure plus the crossing symmetry of the Boltzmann weights.

The face version of the discrete functional equations is given by  \cite{FRAHM2021},
\eq
D_n(\lambda_1,\lambda_2,\cdots,\lambda_n-1)= A_n(\lambda_1,\lambda_2,\cdots,\lambda_n)[D_n(\lambda_1,\lambda_2,\cdots,\lambda_n)], \label{qKZ}
\en
under the condition that $\lambda_n=u_k$ for arbitrary $k$ and where the linear operator $A_n$ can be written as,
\bear
&A_n(\lambda_1,\lambda_2,\cdots,\lambda_n)[B]^{\alpha_1\alpha_2\dots\alpha_{n+1}}_{\beta_1\beta_2\dots\beta_{n+1}}=\frac{\delta_{\alpha_1 \beta_1}\delta_{\alpha_{n+1} \beta_{n+1}}}{\prod_{k=1}^n (1-(\lambda_k-\lambda_n)^2)} \times \nonumber \\
&\displaystyle\sum_{\gamma_i,\delta_i=\pm} \delta_{\alpha_n\gamma_{n+1}} \prod_{k=1}^{n-1}\weight[\alpha_k][\alpha_{k+1}][\gamma_{k+1}][\gamma_{k}][\lambda_n-\lambda_k] B^{\gamma_1\dots\gamma_{n+1}}_{\delta_1\dots\delta_{n+1}} \\
&\prod_{k=1}^{n-1}\weight[\delta_{k+1}][\beta_{k+1}][\beta_{k}][\delta_{k}][\lambda_k-\lambda_n] \weight[\delta_{n+1}][\beta_{n+1}][\beta_{n}][\delta_{n}][-1]. \nonumber
\ear
By direct inspection, we verified for finite lattices $L=4,8$ that the discrete functional equation (\ref{qKZ}) is satisfied.

In the next section, we are going to propose a suitable ansatz for the density matrix which allows for the  solution of the functional equation for short-distances $n=2,3$ and $4$.

\section{Computation of the reduced density matrix}\label{res}

Due to the fact that the IRF model (\ref{wIRF6V}) is made of two copies of the six-vertex model, and likewise the associated spin chain (\ref{Hirf}) is made of two coupled Heisenberg spin chains,
it is expected that this structure carries over to the density matrix as well. Actually, we realized that the IRF reduced density matrix in the ground state (\ref{density}) can be written, after a trivial reordering of the basis states, as a direct sum of two copies of the density matrix of the Heisenberg spin chain,
\eq
D_n^{\text{IRF}}(\lambda_1,\lambda_2,\dots,\lambda_n)= \frac{1}{2} D_n^{XXX}(\lambda_1,\lambda_2,\dots,\lambda_n) \oplus  \frac{1}{2} D_n^{XXX}(\lambda_1,\lambda_2,\dots,\lambda_n),
\en
Note that since $\tr[D_n^{XXX}(\lambda_1,\lambda_2,\dots,\lambda_n)]=1$, it naturally guarantees the normalization of the IRF density matrix, i.e. $\tr[D_n^{\text{IRF}}(\lambda_1,\lambda_2,\dots,\lambda_n)]=1$.

\subsection{Computation of the two-site density matrix}\label{2siteDM}

The two-site density matrix of the isotropic Heisenberg  model, which originally was written in a vector basis \cite{BJMST}, can be conveniently written in term of projector operators or the identity and permutation matrices, such that
\eq
D_2^{XXX}(\lambda_1,\lambda_2)=\left(\frac{1}{4}-\frac{\omega(\lambda_1,\lambda_2)}{6}\right) I_4 + \frac{\omega(\lambda_1,\lambda_2)}{3} P_{12},
\label{D2XXX}
\en
where $I_4$ is the $4\times4$ identity and $P_{12}$ is the permutation operator,
\eq
P_{12}=\left(\begin{array}{cccc}
1 & 0 & 0 & 0\\
0 & 0 & 1 & 0\\
0 & 1 & 0 & 0\\
0 & 0 & 0 & 1\\
\end{array} \right).
\en

Choosing
\eq
\{ \ket{+++},\ket{++-},\ket{+--},\ket{+-+} \} \cup \{\ket{-+-},\ket{-++},\ket{--+},\ket{---} \},
\en
as the basis ordering, the two-site IRF density matrix can be written as,
\eq
D_2^{\text{IRF}}(\lambda_1,\lambda_2)= \frac{1}{2} D_2^{XXX}(\lambda_1,\lambda_2) \oplus  \frac{1}{2} D_2^{XXX}(\lambda_1,\lambda_2).
\label{D2IRF}
\en

On the quantum spin chain side, one has to take the partial trace of the $D_2^{\text{IRF}}(\lambda_1,\lambda_2)$, resulting in the one-site density matrix for the Hamiltonian (\ref{Hirf}),
\eq
{\mathbb D}_1(\lambda_1,\lambda_2)=\left( \begin{array}{cc}
	\frac{1}{2} & \frac{\omega(\lambda_1,\lambda_2)}{3} \\
	\frac{\omega(\lambda_1,\lambda_2)}{3} & \frac{1}{2}
\end{array}\right).
\en

By replacing (\ref{D2XXX}) into (\ref{D2IRF}) and substituting this ansatz in the functional equation (\ref{qKZ}), we obtain a single discrete functional equation for the function $\omega(\lambda_1,\lambda_2)$, which reads,
\eq
\omega(\lambda_1, \lambda_2-1) + \frac{(\lambda_1 - \lambda_2) (\lambda_1 - \lambda_2 + 2)}{(\lambda_1 - \lambda_2)^2-1} \omega(\lambda_1, \lambda_2) = \frac{3}{2}\frac{1}{ (\lambda_1 - \lambda_2)^2-1},
\label{D2qKZ}
\en
for $\lambda_2=u_k$ for $k=1,2,\dots,L$. This discrete functional equation is the same one obtained for the isotropic Heisenberg spin chain \cite{BK2012}. We have verified that this equation is fulfilled by direct inspection for small lattice sizes $L=4,8$.

The only non-trivial one-point correlation function is given by
\eq
\langle \sigma_i^x \rangle_{L} = \frac{2}{3} \omega(0,0).
\label{X}
\en

\subsubsection{Solution in the thermodynamical limit ($L\to \infty$)}

In the thermodynamic limit, there will be arbitrarily many $u_k$ forming a continuum, which allows the equation (\ref{D2qKZ}) to hold for arbitrary values of $\lambda_1$ and $\lambda_2$. Therefore, in the thermodynamic limit, we may remove the restriction on the $\lambda_2$ variable and (\ref{D2qKZ}) becomes an equation for the difference of the variables $\lambda=\lambda_1-\lambda_2$ such that $\omega(\lambda_1,\lambda_2)=\omega_{\infty}(\lambda_1-\lambda_2)$,
\eq
\omega_{\infty}(\lambda+1) + \frac{\lambda (\lambda+ 2)}{\lambda^2-1} \omega_{\infty}(\lambda) = \frac{3}{2}\frac{1}{ \lambda^2-1},
\en
which is exactly the same equation in \cite{BJMST}. The solution in the thermodynamical limit can be written as\cite{BJMST},
\eq
\omega_{\infty}(\lambda)=(\lambda^2-1)\frac{d}{d\lambda}\log\left\{\frac{\Gamma(1+\frac{\lambda}{2})\Gamma(\frac{1}{2}-\frac{\lambda}{2})}{\Gamma(1-\frac{\lambda}{2})\Gamma(\frac{1}{2}+\frac{\lambda}{2})} \right\}+\frac{1}{2}.
\label{omegainf}
\en

Taking the homogeneous limit $\lambda_k=\lambda= 0$, we obtain that
\eq
\omega_{\infty}(0)=\frac{1}{2}-2\log(2),
\en
which implies that for $L\to \infty$,
\eq
\langle \sigma_i^x \rangle_{\infty}=\frac{1}{3}-\frac{4}{3}\log(2)= -0.590862907413...
\en

\subsection{Computation of the three-site density matrix}\label{3siteDM}

Again, the three-site density matrix of the Heisenberg  model can be conveniently written in terms of  identity and permutation matrices acting on three different sites, such that
\eq
D_3^{XXX}(\lambda_1,\lambda_2,\lambda_3)=\rho_1^{(3)} I_8 + \rho_2^{(3)} P_{12}  +\rho_3^{(3)} P_{23} + \rho_4^{(3)} P_{23}P_{12} + \rho_5^{(3)} P_{12}P_{23},
\label{D3XXX}
\en
where $\rho_k^{(3)}=\rho_k^{(3)}(\lambda_1,\lambda_2,\lambda_3)$ are functions that are determined by normalization of the density matrix, by the functional equations and by the density matrix symmetries \cite{BJMST,BST}, which can be written as,
\bear
\rho_1^{(3)}\!&=&\!\tfrac{1}{8} \!-\!\tfrac{1}{12}\!\left(\!1\!-\!\tfrac{1}{\lambda_{13}\lambda_{23}}\! \right)\! \omega(\lambda_{1},\!\lambda_2) \!+\!
\tfrac{1}{12}\!\left(\!1-\!\tfrac{1}{\lambda_{12}\lambda_{23}}\! \right)\! \omega(\lambda_1,\!\lambda_3) \!-\! \tfrac{1}{12}\!\left(\!1\!-\!\tfrac{1}{\lambda_{12}\lambda_{13}} \!\right)\! \omega(\lambda_2,\! \lambda_3), \nonumber \\
\rho_2^{(3)}\!&=&\!\tfrac{1}{6}\!\left(\!1\!-\!\tfrac{1}{\lambda_{13}\lambda_{23}}\! \right)\! \omega(\lambda_{1},\!\lambda_2) \!-\!
\tfrac{1}{6}\!\left(\!1-\!\tfrac{1}{\lambda_{12}\lambda_{23}}\! \right)\! \omega(\lambda_1,\!\lambda_3) \!-\! \tfrac{1}{6}\!\left(\!\tfrac{1}{\lambda_{12}\lambda_{13}} \!\right)\! \omega(\lambda_2,\!\lambda_3), \nonumber\\
\rho_3^{(3)}\!&=&\!\!\!-\tfrac{1}{6}\!\left(\!\tfrac{1}{\lambda_{13}\lambda_{23}}\! \right)\! \omega(\lambda_{1},\!\lambda_2) \!-\!
\tfrac{1}{6}\!\left(\!1-\!\tfrac{1}{\lambda_{12}\lambda_{23}}\! \right)\! \omega(\lambda_1,\!\lambda_3) \!+\! \tfrac{1}{6}\!\left(\!1\!-\!\tfrac{1}{\lambda_{12}\lambda_{13}} \!\right)\! \omega(\lambda_2,\! \lambda_3), \\
\rho_4^{(3)}\!&=&\!\!\!\tfrac{1}{12}\!\!\left(\!\tfrac{2-\lambda_{12}}{\lambda_{13}\lambda_{23}}\! \right)\! \omega(\lambda_{1},\!\lambda_2) \!+\!
\tfrac{1}{12}\!\!\left(\!2-\!\tfrac{2}{\lambda_{12}\lambda_{23}}\! -\!\tfrac{1}{\lambda_{12}}\!+\!\tfrac{1}{\lambda_{23}}\right)\! \omega(\lambda_1,\!\lambda_3) \nonumber\\
&+&\! \tfrac{1}{12}\!\!\left(\!\tfrac{2}{\lambda_{12}\lambda_{13}}\!+\!\tfrac{1}{\lambda_{12}}\!-\! \tfrac{1}{\lambda_{13}} \!\right)\! \omega(\lambda_2,\! \lambda_3), \nonumber \\
\rho_5^{(3)}\!&=&\!\!\!
\tfrac{1}{12}\!\!\left(\!\tfrac{2}{\lambda_{12}\lambda_{23}}-\!\tfrac{2}{\lambda_{12}\lambda_{23}}\! -\!\tfrac{1}{\lambda_{13}}\!+\!\tfrac{1}{\lambda_{23}}\right)\! \omega(\lambda_{1},\!\lambda_2) \!+\!
\tfrac{1}{12}\!\!\left(\!2-\!\tfrac{2}{\lambda_{12}\lambda_{23}}\! +\!\tfrac{1}{\lambda_{12}}\!-\!\tfrac{1}{\lambda_{23}}\right)\! \omega(\lambda_1,\!\lambda_3) \nonumber \\
&+&\! \tfrac{1}{12}\!\!\left(\!\tfrac{2}{\lambda_{12}\lambda_{13}}\!-\!\tfrac{1}{\lambda_{12}}\!+\! \tfrac{1}{\lambda_{13}} \!\right)\! \omega(\lambda_2,\! \lambda_3), \nonumber
\ear
where $\lambda_{ij}=\lambda_i-\lambda_j$.

Choosing
\begin{align}
&\!\!\{ \ket{+\!+\!++}\!,\!\ket{+\!+\!+-}\!,\!\ket{+\!+\!--}\!,\!\ket{+\!+\!-+}\!,\!\ket{+\!-\!--}\!,\!\ket{+\!-\!-+}\!,\!\ket{+\!-\!++}\!,\!\ket{+\!-\!+-}\!\}  \\
& \!\!\!\cup\! \{ \ket{-\!+\!-+}\!,\!\ket{-\!+\!--}\!,\!\ket{-\!+\!+-}\!,\! \ket{-\!+\!++}\!,\!\ket{-\!-\!+-}\!,\!\ket{-\!-\!++}\!,\!\ket{-\!-\!-+}\!,\!\ket{-\!-\!--}\!\}, \nonumber
\end{align}
as the basis ordering, the three-site IRF density matrix can be written as,
\eq
D_3^{\text{IRF}}(\lambda_1,\lambda_2,\lambda_3)= \tfrac{1}{2} D_3^{XXX}(\lambda_1,\lambda_2,\lambda_3) \oplus  \tfrac{1}{2} D_3^{XXX}(\lambda_1,\lambda_2,\lambda_3).
\label{D3IRF}
\en

On the quantum spin chain side, one has to take the partial trace on  $D_3^{\text{IRF}}(\lambda_1,\lambda_2,\lambda_3)$, which results in the two-site density matrix of the Hamiltonian (\ref{Hirf}),
\eq
{\mathbb D}_2(\lambda_1,\lambda_2,\lambda_3)=\left( \begin{array}{cccc}
	\tfrac{1}{4} & \tfrac{\omega(\lambda_2,\lambda_3)}{6} & \tfrac{\omega(\lambda_1,\lambda_2)}{6} & \tfrac{\Omega^{(3)}(\lambda_1,\lambda_2,\lambda_3)}{6}	\\
	\tfrac{\omega(\lambda_2,\lambda_3)}{6} & \tfrac{1}{4}  &  \tfrac{\Omega^{(3)}(\lambda_1,\lambda_2,\lambda_3)}{6} & \tfrac{\omega(\lambda_1,\lambda_2)}{6} \\
	\tfrac{\omega(\lambda_1,\lambda_2)}{6} &  \tfrac{\Omega^{(3)}(\lambda_1,\lambda_2,\lambda_3)}{6} & \tfrac{1}{4}  & \tfrac{\omega(\lambda_2,\lambda_3)}{6} \\
	\tfrac{\Omega^{(3)}(\lambda_1,\lambda_2,\lambda_3)}{6} &  \tfrac{\omega(\lambda_1,\lambda_2)}{6} & \tfrac{\omega(\lambda_2,\lambda_3)}{6} & \tfrac{1}{4} 	
\end{array}\right),
\en
where
\eq
\Omega^{(3)}(\lambda_1,\lambda_2,\lambda_3)=3\left(\rho_4^{(3)}+\rho_5^{(3)}\right)= \tfrac{\omega(\lambda_1,\lambda_2)}{ \lambda_{13}\lambda_{23}} +\omega(\lambda_1,\lambda_3)\left(1 - \tfrac{1}{\lambda_{12}\lambda_{23}} \right) + \tfrac{\omega(\lambda_2,\lambda_3)}{\lambda_{12}\lambda_{13}},
\en

The only non-trivial two-site correlation function is obtained from $\Omega^{(3)}(\lambda_1,\lambda_2,\lambda_3)$ as,
\bear
\langle \sigma_i^x \sigma_{i+1}^x   \rangle_{L}&=& \tfrac{2}{3} \Omega^{(3)}(0,0,0),
\label{XX12}
\ear
where the homogeneous limit is a singular one, which results in $\Omega^{(3)}(0,0,0)=\omega^{(0,0)} + \omega^{(1,1)} - \omega^{(2,0)}/2$, where $\omega^{(m,n)}=\partial_{\lambda_1}^m \partial_{\lambda_2}^n \omega(\lambda_1,\lambda_2)|_{\lambda_1=\lambda_2=0}$ for arbitrary system size $L$.

The homogeneous limit of $\Omega(\lambda_1,\lambda_2,\lambda_3)$ in the thermodynamical limit gives,
\eq
\Omega^{(3)}_{\infty}(0,0,0)=\tfrac{1}{2} - 8 \log(2) + \tfrac{9}{2} \zeta(3),
\en
therefore,
\bear
\langle \sigma_1^x \sigma_2^x   \rangle_{\infty}&=& \tfrac{1}{3} - \tfrac{16}{3}\log(2) + 3 \zeta(3)=0.242719079825...
\ear

\subsection{Computation of the four-site density matrix}\label{4siteDM}

The four-site density matrix of the Heisenberg model was computed in \cite{BST} and can be written as,
\bear
D_4^{XXX}(\lambda_1,\lambda_2,\lambda_3,\lambda_4)=\sum_{k=1}^{14}\rho_k^{(4)} \check{P}_k,
\ear
where $\check{P}_k$ for $k=1,\dots,14$ can be taken from the following ordered set of linearly independent operators,
\bear
&&\{\check{P}_k\}_{k=1}^{14}=\{ I_{16}, ~P_{12}, ~P_{23}, ~P_{34}, ~P_{12}P_{23}, ~P_{23}P_{12}, ~P_{23}P_{34}, ~P_{34}P_{23}, ~P_{12}P_{34}, ~P_{13}P_{24},
\nonumber \\
&& ~P_{12}P_{34} P_{23}, ~P_{12}P_{23} P_{34},  ~P_{34}P_{23} P_{12}, ~P_{23}P_{34} P_{12}  \}, \nonumber
\ear
and the coefficients $\rho_k^{(4)}=\rho_k^{(4)}(\lambda_1,\lambda_2,\lambda_3,\lambda_4)$, which can be read from \cite{BST}\footnote{Note that it was used in \cite{BST} a function $G$, which is closely related to $\omega$, such as $G(x)=\omega(\im x)-\tfrac{1}{2}$.}, have the following structure (we list the coefficients $A_i^{(k)}$ and $B_i^{(k)}$ in the Appendix A),
\begin{align}
	\rho^{(4)}_k = p_0^{(k)} &+ A_{1}^{(k)}(\lambda_1,\lambda_2,\lambda_3,\lambda_4)  ~\omega(\lambda_1,\lambda_2) +A_{2}^{(k)}(\lambda_1,\lambda_2,\lambda_3,\lambda_4) ~ \omega(\lambda_1,\lambda_3) \nonumber\\
	&+ A_{3}^{(k)}(\lambda_1,\lambda_2,\lambda_3,\lambda_4) ~ \omega(\lambda_1,\lambda_4) + A_{4}^{(k)}(\lambda_1,\lambda_2,\lambda_3,\lambda_4) ~  \omega(\lambda_2,\lambda_3) \nonumber \\
	&+ A_{5}^{(k)}(\lambda_1,\lambda_2,\lambda_3,\lambda_4) ~ \omega(\lambda_2,\lambda_4) +A_{6}^{(k)}(\lambda_1,\lambda_2,\lambda_3,\lambda_4) ~ \omega(\lambda_3,\lambda_4) \label{RHO4}\\
	&+B_{1}^{(k)}(\lambda_1,\lambda_2,\lambda_3,\lambda_4) ~ \omega(\lambda_1,\lambda_2) ~ \omega(\lambda_3,\lambda_4) \nonumber \\
	&+ B_{2}^{(k)}(\lambda_1,\lambda_2,\lambda_3,\lambda_4) ~ \omega(\lambda_1,\lambda_3)~\omega(\lambda_2,\lambda_4) \nonumber \\
	&+ B_{3}^{(k)}(\lambda_1,\lambda_2,\lambda_3,\lambda_4) ~ \omega(\lambda_1,\lambda_4)~\omega(\lambda_2,\lambda_3). \nonumber
\end{align}
Due to the factorization properties of the correlation functions for the Heisenberg chain \cite{BJMST,BST}, the functions $\rho_k^{(3)}$ and $\rho_k^{(4)}$ can be written in terms of the two-site functions $\omega(\lambda_i,\lambda_j)$ and likewise all non-trivial correlations below.

The non-trivial correlations can be expressed in terms of the homogeneous limit of the following functions,
\bear
\Omega_1^{(4)}(\lambda_1,\lambda_2,\lambda_3,\lambda_4)&=& 2 \left(\rho^{(4)}_{11} +  \rho^{(4)}_{12} +  \rho^{(4)}_{13} +  \rho^{(4)}_{14}\right),  \\
\Omega_2^{(4)}(\lambda_1,\lambda_2,\lambda_3,\lambda_4)&=&4 \left(\rho^{(4)}_9 +  \rho^{(4)}_{10} \right) + \Omega_1^{(4)}(\lambda_1,\lambda_2,\lambda_3,\lambda_4),  \nonumber \\
\Omega_3^{(4)}(\lambda_1,\lambda_2,\lambda_3,\lambda_4)&=& 2\left(- \rho^{(4)}_{11} +  \rho^{(4)}_{12} +  \rho^{(4)}_{13} -  \rho^{(4)}_{14}\right),  \nonumber \\
\Omega_4^{(4)}(\lambda_1,\lambda_2,\lambda_3,\lambda_4)&=& 8 \rho^{(4)}_{3} + 4 \left( \rho^{(4)}_{5} +  \rho^{(4)}_{6} +  \rho^{(4)}_{7} +  \rho^{(4)}_8\right) + \Omega_1^{(4)}(\lambda_1,\lambda_2,\lambda_3,\lambda_4),  \nonumber
\ear
where the last function $\Omega_4^{(4)}(\lambda_1,\lambda_2,\lambda_3,\lambda_4)$ can be further simplified as $\Omega_4^{(4)}(\lambda_1,\lambda_2,\lambda_3,\lambda_4)=\tfrac{1}{3}\left(-4 ~\omega(\lambda_1,\lambda_2) +6~\omega(\lambda_2,\lambda_3) \right)$.

This implies that the non-trivial three-points correlations are given by,
\bear
\langle \sigma_i^x \sigma_{i+1}^x  \sigma_{i+2}^x   \rangle_L\!&=&\!\! \Omega_1^{(4)}(0,0,0,0), \nonumber \\
\!&=&\!\!\omega^{(0, 0)}\left[\tfrac{2}{3} \!+\! \tfrac{4}{3} \omega^{(1, 1)} \!+\! \tfrac{2}{9}\omega^{(2, 2)} \!-\! \tfrac{4}{27}\omega^{(3, 1)} \right]
\!-\! \omega^{(1, 0)}\!\left[ \tfrac{4}{3}\omega^{(1, 0)} \!+\! \tfrac{4}{9} \omega^{(2, 1)} \!-\! \tfrac{4}{27}\omega^{(3, 0)} \right] \nonumber \\
\!&-&\!\! \tfrac{1}{9} \omega^{(3, 1)} + \left[4\omega^{(1, 1)} - 2\omega^{(2, 0)}\right]
\left[ \tfrac{1}{3} + \tfrac{1}{9}\omega^{(2, 0)}\right]
+ \tfrac{1}{6}\omega^{(2, 2)},  \label{XXX}
\ear
\bear
\langle \sigma_i^x  \sigma_{i+2}^x   \rangle_L&=&\!\! \Omega_2^{(4)}(0,0,0,0), \nonumber\\
&=&\!\! \omega^{(0,0)}\left[\tfrac{4}{5}\omega^{(0,0)}+\tfrac{8}{15}\omega^{(1,1)} + \tfrac{7}{45}\omega^{(2,2)} - \tfrac{14}{135}\omega^{(3,1)}\right]   \nonumber\\
&-&\!\!  \omega^{(1,0)} \left[ \tfrac{8}{15} \omega^{(1,0)} +  \tfrac{14}{45} \omega^{(2,1)} -  \tfrac{14}{135} \omega^{(3,0)} \right] \label{XX13} \\
&+&\!\! \omega^{(1,1)}\left[\tfrac{2}{5} +\tfrac{14}{45} \omega^{(2,0)}\right] -\omega^{(2,0)}\left[\tfrac{4}{15} +\tfrac{7}{45} \omega^{(2,0)}\right]
+ \tfrac{2}{15} \omega^{(2,2)} -  \tfrac{4}{45} \omega^{(3,1)}, \nonumber \\
\langle \sigma_i^y  \sigma_{i+2}^y   \rangle_L &=&\!\! \Omega_3^{(4)}(0,0,0,0),  \nonumber \\
&=&\!\! \omega^{(0,0)}\left[\tfrac{4}{15} \omega^{(0,0)} +\tfrac{2}{5} \omega^{(1,1)}  +\tfrac{4}{45} \omega^{(2,2)} - \tfrac{8}{135} \omega^{(3,1)}    \right] \nonumber \\
&-&\!\!  \omega^{(1,0)} \left[ \tfrac{2}{5} \omega^{(1,0)} +  \tfrac{8}{45} \omega^{(2,1)} -  \tfrac{8}{135} \omega^{(3,0)} \right] + \omega^{(1,1)} \left[ \tfrac{7}{15}  +  \tfrac{8}{45} \omega^{(2,0)} \right] \nonumber \\
&-&\!\! \omega^{(2,0)} \left[ \tfrac{1}{5}  +  \tfrac{4}{45} \omega^{(2,0)} \right] -\tfrac{1}{15} \omega^{(3,1)} + \tfrac{1}{10} \omega^{(2,2)}. \label{YY13}
\ear

The remaining non-zero three-points correlations are related to the previous ones as follows,
\bear
\langle \sigma_i^z  \sigma_{i+2}^z   \rangle_L &=& \Omega_4^{(4)}(0,0,0,0)=\tfrac{2}{3}\omega(0,0)=\langle \sigma_i^x   \rangle_L, \nonumber  \\
\langle \sigma_i^y \sigma_{i+1}^x  \sigma_{i+2}^y   \rangle_L&=& -\langle \sigma_i^y  \sigma_{i+2}^y   \rangle_L, \nonumber \\
\langle \sigma_i^z \sigma_{i+1}^x  \sigma_{i+2}^z   \rangle_L&=& -\langle \sigma_i^z  \sigma_{i+2}^z   \rangle_L.
\ear

In the thermodynamical limit, one can evaluate the correlation functions by use of the function $\omega_{\infty}(\lambda)$ given in (\ref{omegainf}). The final results are given by,
\bear
\langle \sigma_i^x \sigma_{i+1}^x \sigma_{i+2}^x   \rangle_{\infty}\!\!\!\!&=&\!\!\!
\tfrac{1}{3} \!-\! 12 \log(2) \!+\! \tfrac{74}{3} \zeta(3) \!-\! \tfrac{56}{3} \log(2) \zeta(3) \!-\!
6 \zeta(3)^2 \!-\! \tfrac{125}{6} \zeta(5) \!+\! \tfrac{100}{3} \log(2) \zeta(5), \nonumber \\
&=&-0.200994509028...,\nonumber\\
\langle \sigma_i^x  \sigma_{i+2}^x   \rangle_{\infty}\!\!\!&=&\!\!\! \tfrac{1}{5} \!-\! \tfrac{16}{3} \log(2) \!+\! \tfrac{232}{15} \zeta{(3)} \!-\! \tfrac{32}{3} \log(2) \zeta{(3)} \!-\!
\tfrac{21}{5} \zeta{(3)}^2 \!-\! \tfrac{95}{6} \zeta{(5)} \!+\! \tfrac{70}{3} \log(2) \zeta{(5)},\nonumber \\
&=&0.491445392361...,\nonumber\\
\langle \sigma_i^y  \sigma_{i+2}^y   \rangle_{\infty}\!\!\!&=&\!\!\! \tfrac{1}{15} \!-\! 4 \log(2) \!+\! \tfrac{169}{15}\zeta{(3)} \!-\! \tfrac{20}{3} \log(2) \zeta{(3)} \!-\! \tfrac{12}{5} \zeta{(3)}^2 \!-\! \tfrac{65}{6} \zeta{(5)} \!+\! \tfrac{40}{3} \log(2)  \zeta{(5)}, \nonumber  \\
&=&0.164575433372...
\ear

\section{Integral equations for finite system size}\label{NLIEsec}

The physical properties of the IRF six-vertex model and its associated spin chain was obtained from the leading eigenvalue of the six-vertex model transfer matrix with periodic boundary conditions via non-linear integral equations \cite{TAVARES2023}. The same should apply to the case of correlation functions.

More specifically, one can obtain the leading eigenvalue $\Lambda_0(\lambda)$ of the six-vertex row-to-row transfer matrix given by \cite{KLUMPER93},
\eq
\ln\left[\frac{\Lambda_0(\im x-\tfrac{1}{2})}{(\im x+\tfrac{1}{2})^L}\right]= L ~ e(x+\tfrac{\im}{2}) +   \tfrac{\im\pi L}{2}+  \left( K \ast \ln{B \bar{B}}\right)(x), \label{NLIEeigenvalue}
\en
where we set from now on $\lambda=\im x$, 
\eq
e(x)=\log\left[\frac{ \Gamma\left(1- \frac{\im x}{2}\right)\Gamma\left(\frac{1}{2}+ \frac{\im x}{2}\right)}{ \Gamma\left(1+ \frac{\im x}{2}\right)\Gamma\left(\frac{1}{2}- \frac{\im x}{2}\right)}\right],
\en
and
\eq
K(x)=\frac{\pi}{\cosh{ \pi x}}.
\en
The symbol $\ast$ denotes convolution $(f*g)(x)=\frac{1}{2 \pi}\int_{-\infty}^{\infty} f(x-y)g(y)dy$.

The auxiliary functions $b(x)$, $\bar{b}(x)$ and its simply related functions
$B(x)=b(x)+1$ and $\bar{B}(x)=\bar{b}(x)+1$ are solutions of the following set
of non-linear integral equations \cite{KLUMPER92},
\begin{subequations}
\begin{align}
\ln{b(x)}&=  L \ln(\tanh(\tfrac{\pi x}{2}))  +  \left( \!F\!\ast\! \ln{B}\right)\!(x) -
\left(\! F\!\ast\! \ln{\bar{B}}\right)\!(x+\im), \label{NLIE1}\\
\ln{\bar{b}(x)}&= L \ln(\tanh(\tfrac{\pi x}{2}))  - \left(\! F \!\ast\! \ln{B}\right)\!(x-\im) + \left(\! F\!\ast\! \ln{\bar{B}}\right)\!(x), \label{NLIE2}
\end{align}
\end{subequations}
where
the Kernel function is given by
\eq
F(x)=\im \frac{d e(x) }{dx}=\int_{-\infty}^{\infty}\frac{e^{-|k|/2+\im k x}}{2 \cosh(\tfrac{k}{2})} dk.
\en

Similarly, the two-site correlation function $\omega(\lambda_1,\lambda_2)$ is given by \cite{CORRXXZ},
\eq
\omega(\lambda_1,\lambda_2)=\left(\lambda_{12}^2 - 1\right)\frac{\Psi(\im \lambda_1+\tfrac{\im}{2},\im\lambda_2+\tfrac{\im}{2})}{2}+\frac{1}{2},
\en
where
\eq
\Psi(\im \lambda_1+\tfrac{\im}{2},\!\im\lambda_2+\tfrac{\im}{2})\!=\!2 F(\lambda_1 -\!\lambda_2)+\! \int_{-\infty}^{\infty}\!\!\frac{1}{\cosh(\pi(\lambda_2\!+\!\frac{\im}{2}\!-\!x))} \!\left[\! \frac{g_{\lambda_1}^{(+)}(x)}{1+b^{-1}(x)}\!+\! \frac{g_{\lambda_1}^{(-)}(x)}{1+\bar{b}^{-1}(x)} \!\right]\!dx,
\en
and the additional auxiliary functions $g_{\lambda_1}^{(\pm)}(x)$ are solution of the following set of linear integral equations \cite{CORRXXZ},
\begin{subequations}
\begin{align}
g_{\lambda_1}^{(+)}(x)&=   \frac{\pi}{\cosh(\pi(\lambda_1+\tfrac{\im}{2}-x))}  +  \left( \!F\!\ast\! \frac{g_{\lambda_1}^{(+)}}{1+b^{-1}} \right)\!(x) -
\left(\! F\!\ast\! \frac{g_{\lambda_1}^{(-)}}{1+\bar{b}^{-1}}\right)\!(x+\im), \label{NLIEG1}\\
g_{\lambda_1}^{(-)}(x)&=   \frac{\pi}{\cosh(\pi(\lambda_1+\tfrac{\im}{2}-x))}  -  \left( \!F\!\ast\! \frac{g_{\lambda_1}^{(+)}}{1+b^{-1}} \right)\!(x-\im) +
\left(\! F\!\ast\! \frac{g_{\lambda_1}^{(-)}}{1+\bar{b}^{-1}}\right)\!(x).
\label{NLIEG2}
\end{align}
\end{subequations}

\begin{table}[ht]
	\begin{center}
		\begin{tabular}{|l|c|c|c|c|c|}
			\hline
			Length & $\langle \sigma_1^x \rangle$  & $\langle \sigma_1^x \sigma_2^x \rangle$ & $\langle \sigma_1^x \sigma_2^x  \sigma_3^x \rangle$ & $\langle \sigma_1^x \sigma_3^x \rangle$ & $\langle \sigma_1^y \sigma_3^y \rangle$ \\
			\hline
			$L=4$ & $-0.66666667$ &  $0.33333333$ & $-0.66666667$ & $1.00000000$  & $0.66666667$ \\
			\hline
			$L=8$ & $-0.60851556$ &  $0.26103720$ & $-0.25193710$ & $0.55630211$ & $0.21746487$ \\
			\hline
			$L=12$ &$-0.59859899$ &  $0.25044371$ & $-0.22109565$ & $0.51802986$ & $0.18542814$ \\
			\hline
			$L=16$ &$-0.59519136$ &  $0.24696584$ & $-0.21183645$ & $0.50601523$ & $0.17583391$ \\
			\hline
			$L=32$ & $-0.59193864$ & $0.24374937$ &	$-0.20358916$ &	$0.49500263$ & $0.16727766$ \\
			\hline			
			$L=64$ & $-0.59113127$ & $0.24297329$ & $-0.20163433$ & $0.49232982$ & $0.16524315$ \\
			\hline
			$L=128$ & $-0.59092994$ & $0.24278223$ & $-0.20115366$ & $0.49166622$ &	$0.16474172$ \\
			\hline
			$L=256$	& $-0.59087965$ & $0.24273481$ & $-0.20103420$ & $0.49150058$ & $0.16461694$ \\
			\hline
			$L=512$ & $-0.59086709$ & $0.24272301$ & $-0.20100442$ & $0.49145918$ & $0.16458580$ \\		
			\hline
			$L=1024$ &$-0.59086395$ &  $0.24272006$ & $-0.20099698$ & $0.49144884$ & $0.16457802$ \\
			\hline
			$L\rightarrow\infty$ & $-0.59086290$ & $0.24271907$ & $-0.20099450$ & $0.49144539$  & $0.16457543$ \\
			\hline
		\end{tabular}
	\end{center}
	\caption{Comparison of numerical results from exact diagonalization for $L=4,8,12$ sites and from the solution of the set of integral equations  with the analytical result in the thermodynamic limit for correlations.}
	\label{table}
\end{table}

\begin{figure}[thb]
	\begin{center}
		\includegraphics[width=0.65\linewidth, angle=-90]{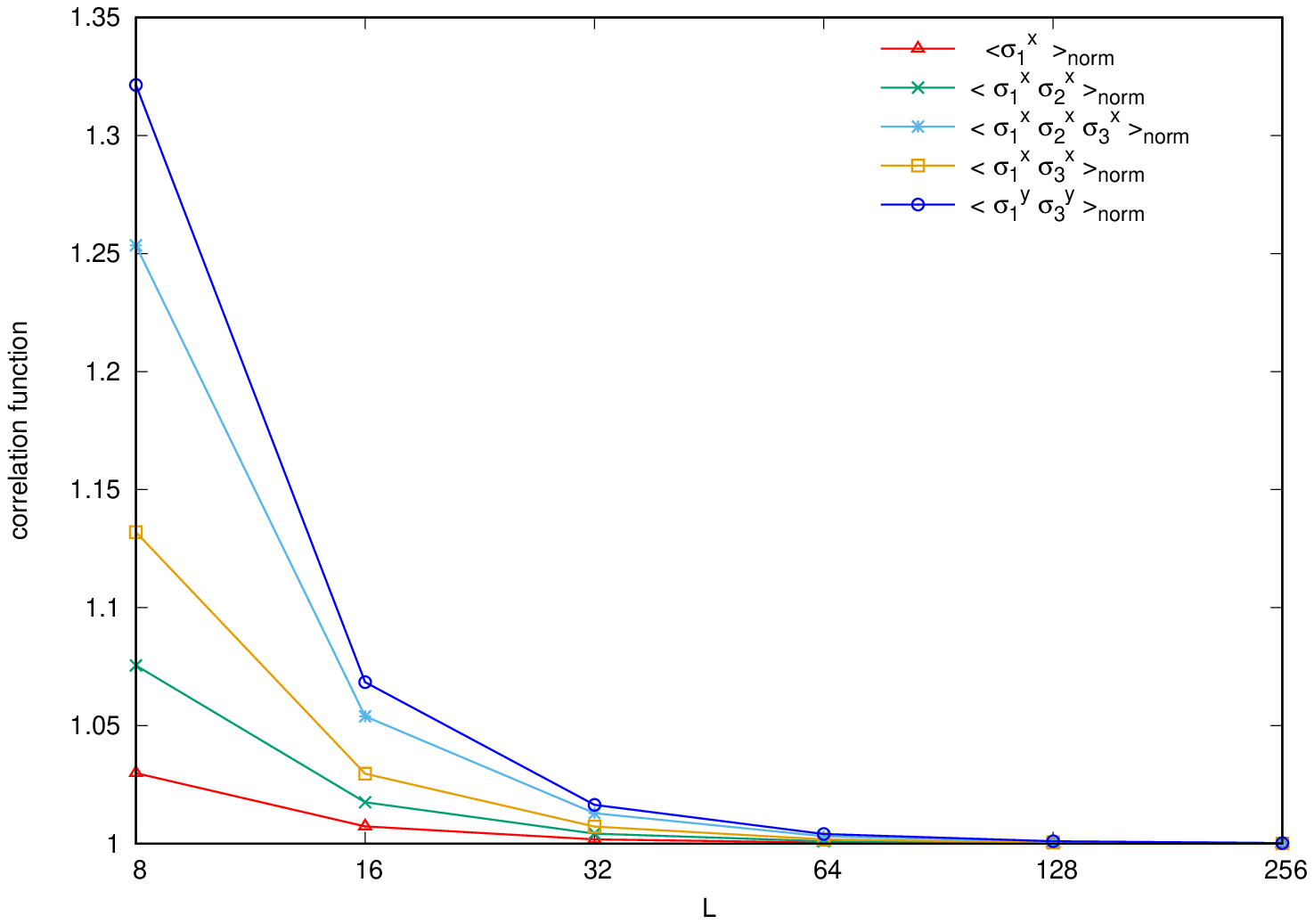}
		\caption{Correlations of the six-vertex IRF spin chain (\ref{Hirf}) normalized by its values at the thermodynamic limit (e.g $\langle \sigma_1^x \rangle_{\text{norm}}=\langle \sigma_1^x \rangle_L/\langle \sigma_1^x \rangle_{\infty}$) versus system size.}
		\label{fig}
	\end{center}
\end{figure}

By iteratively solving the equations (\ref{NLIE1},\ref{NLIE2}) and (\ref{NLIEG1},\ref{NLIEG2}), one can evaluate  $\omega(0,0)$  and its derivatives $\omega^{(n,m)}$ for different values of the system size $L$. This allows us to compute the correlation function given in Eq.(\ref{X}), Eq.(\ref{XX12}) and Eqs.(\ref{XXX}-\ref{YY13}). The results are listed in the Table \ref{table}. They are in good agreement with the exact diagonalization at smaller lattice sizes $L=4,8,12,$ as well as  with the thermodynamic limit values. The results are also shown in the Figure \ref{fig}.

It is worth mentioning that, due to the inherent relationship between the spin chains, the correlations of the IRF spin chain in the ground state  ended up having the same values as the Heisenberg (XXX) spin chain, which are $\langle \sigma_1^x \rangle_{\text{IRF}}=\langle \sigma_1^z \sigma_2^z \rangle_{XXX}$,  $\langle \sigma_1^x \sigma_2^x \rangle_{\text{IRF}}=\langle \sigma_1^z \sigma_3^z \rangle_{XXX}$, $\langle \sigma_1^x \sigma_2^x  \sigma_3^x \rangle_{\text{IRF}}=\langle \sigma_1^z \sigma_4^z \rangle_{XXX}$, $\langle \sigma_1^x \sigma_3^x   \rangle_{\text{IRF}}=\langle \sigma_1^z \sigma_2^z \sigma_3^z \sigma_4^z \rangle_{XXX}$
and $\langle \sigma_1^y \sigma_3^y   \rangle_{\text{IRF}}=\langle \sigma_1^z \sigma_2^x \sigma_3^x \sigma_4^z \rangle_{XXX}$.

\section{Conclusion}\label{CONCLUSION}

We computed short-distance correlations of the IRF version of the isotropic six-vertex model and its associated spin chain. This was done by exploiting the fact that the density matrix of the model satisfies the discrete version of the qKZ equation. By proposing a suitable ansatz for the density matrix, we could exploit the close relationship between the IRF spin chain and the Heisenberg spin chain. We computed explicitly the $n=2,3,4$ sites correlations for finite system sizes via non-linear integral equations and in the thermodynamic limit. The results show very good agreement.

Numerical equality of the different correlators of the Heisenberg spin chain and its IRF counterpart should not be seen as a trivial fact, as the correct identification between these is not a given. It draws from the recent construction of the density matrix for  IRF models\cite{FRAHM2021,FRAHM2023}. Besides, it is worth to emphasize that this identification occurs for the correlators evaluated in the ground state.
Furthermore, we had already pointed it out how the different order sets in either model, despite the same integrable structure which is a consequence of the mapping from one onto the other\cite{TAVARES2023}.

We expect that, far from being an isolate example, new quantum spin chains can be obtained from other IRF models related to more general vertex models and that the correlation functions can be evaluated via functional equations of the quantum Knizhnik-Zamolodchikov.  We hope to report on these problems in the future.

\section*{Acknowledgments}

GAPR thanks FAPESP (grant number 2023/03947-0) for funding.

\newpage

\section*{\bf Appendix A: Coefficients of $\rho_k^{(4)}(\lambda_1,\lambda_2,\lambda_3,\lambda_4)$}
\setcounter{equation}{0}
\renewcommand{\theequation}{A.\arabic{equation}}

In this appendix, we list the necessary coefficients $A_i^{(k)}$ and $B_i^{(k)}$ needed in Eq.(\ref{RHO4}) in order to compute the correlations given by the functions $\Omega_i(\lambda_1,\lambda_2,\lambda_3,\lambda_4)$
for $i=1,2,3$. The constant term is given by $p_0^{(1)}=1/16$ and $p_0^{(k)}=0$ for $k>1$.
We can conveniently write $A_1^{(k)}(\lambda_1,\lambda_2,\lambda_3,\lambda_4)$ and  $B_1^{(k)}(\lambda_1,\lambda_2,\lambda_3,\lambda_4)$ as,
\bear
A_1^{(k)}(\lambda_1,\lambda_2,\lambda_3,\lambda_4)&=&\frac{Q_1^{(k)}(\lambda_1,\lambda_2,\lambda_3,\lambda_4)}{\lambda_{13} \lambda_{14}\lambda_{23}\lambda_{24}}, \nonumber \\
B_1^{(k)}(\lambda_1,\lambda_2,\lambda_3,\lambda_4)&=&\frac{Q_2^{(k)}(\lambda_1,\lambda_2,\lambda_3,\lambda_4)}{\lambda_{13} \lambda_{14}\lambda_{23}\lambda_{24}},
\ear
and the remaining coefficients are given as follows,
\begin{align}
&Q_1^{(9)}(\lambda_1,\lambda_2,\lambda_3,\lambda_4)=-\tfrac{1}{60}\left(14 -\lambda_{12}^2 + 10 \lambda_{13}\lambda_{23} \right),\nonumber \\
&Q_2^{(9)}(\lambda_1,\lambda_2,\lambda_3,\lambda_4)=-\tfrac{ \lambda_{14}\lambda_{24}}{90}\left( 2 - 3 \lambda_{12}^2 - 10 \lambda_{13} \lambda_{23}\right) \nonumber \\
&+ \tfrac{1}{90}\tfrac{\lambda_{24}}{\lambda_{12} }\left( 22 + 2 \lambda_{23}^2 - 6 \lambda_{13} \lambda_{12} - 3 \lambda_{13}^2 \lambda_{12}^2 \right) + \tfrac{1}{90}\tfrac{ \lambda_{14}}{\lambda_{12} }\left(22 + 2 \lambda_{13}^2 + 6 \lambda_{23} \lambda_{12} - 3 \lambda_{23}^2 \lambda_{12}^2  \right), \nonumber\\
&A_2^{(9)}(\lambda_1,\lambda_2,\lambda_3,\lambda_4)=A_1^{(9)}(\lambda_1,\lambda_3,\lambda_2,\lambda_4)-\tfrac{1}{6}, \nonumber \\
&A_3^{(9)}(\lambda_1,\lambda_2,\lambda_3,\lambda_4)=A_1^{(9)}(\lambda_1,\lambda_4,\lambda_3,\lambda_2)-\tfrac{1}{6} \left(1-\tfrac{1}{\lambda_{13}\lambda_{34}}\right), \\
&A_4^{(9)}(\lambda_1,\lambda_2,\lambda_3,\lambda_4)=A_1^{(9)}(\lambda_3,\lambda_2,\lambda_1,\lambda_4)-\tfrac{1}{6}\left(1-\tfrac{1}{\lambda_{23}\lambda_{24}}+\tfrac{1}{\lambda_{23}\lambda_{34}}\right), \nonumber\\
&A_5^{(9)}(\lambda_1,\lambda_2,\lambda_3,\lambda_4)=A_1^{(9)}(\lambda_4,\lambda_2,\lambda_3,\lambda_1), \qquad A_6^{(9)}(\lambda_1,\lambda_2,\lambda_3,\lambda_4)=A_1^{(9)}(\lambda_4,\lambda_3,\lambda_2,\lambda_1), \nonumber\\
&B_2^{(9)}(\lambda_1,\lambda_2,\lambda_3,\lambda_4)=B_1^{(9)}(\lambda_1,\lambda_3,\lambda_2,\lambda_4)-\tfrac{1}{18  \lambda_{12}\lambda_{34}}\left(2 - \lambda_{14} \lambda_{23} +  \lambda_{12}  \lambda_{34}   \right), \nonumber\\
&B_3^{(9)}(\lambda_1,\lambda_2,\lambda_3,\lambda_4)=B_1^{(9)}(\lambda_1,\lambda_4,\lambda_3,\lambda_2) +\tfrac{1}{18  \lambda_{12}\lambda_{34}} \left(2 - \lambda_{13} \lambda_{24} -  \lambda_{12}  \lambda_{34}  \right), \nonumber
\end{align}
\begin{align}
	&Q_1^{(10)}(\lambda_1,\lambda_2,\lambda_3,\lambda_4)=(-\tfrac{1}{6} + \tfrac{ \lambda_{12}^2 -4}{20 \lambda_{12} \lambda_{13} \lambda_{23} \lambda_{14}} - \tfrac{ \lambda_{12}^2 -4}{20 \lambda_{12} \lambda_{13} \lambda_{23} \lambda_{24} })\lambda_{13} \lambda_{14}\lambda_{23}\lambda_{24}, \nonumber\\
	&Q_2^{(10)}(\lambda_1,\lambda_2,\lambda_3,\lambda_4)=\tfrac{1}{90}(  \lambda_{12}^2 -4) ( \lambda_{34}^2-4), ~ A_2^{(10)}(\lambda_1,\lambda_2,\lambda_3,\lambda_4)=A_1^{(10)}(\lambda_1,\lambda_3,\lambda_2,\lambda_4)+\tfrac{1}{6},  \nonumber \\
	&A_3^{(10)}(\lambda_1,\lambda_2,\lambda_3,\lambda_4)=A_1^{(10)}(\lambda_1,\lambda_4,\lambda_3,\lambda_2)+\tfrac{1}{6},   ~ A_4^{(10)}(\lambda_1,\lambda_2,\lambda_3,\lambda_4)=A_1^{(10)}(\lambda_3,\lambda_2,\lambda_1,\lambda_4)+\tfrac{1}{3}, \nonumber\\
	&A_5^{(10)}(\lambda_1,\lambda_2,\lambda_3,\lambda_4)=A_1^{(10)}(\lambda_4,\lambda_2,\lambda_3,\lambda_1) +\tfrac{1}{6}, ~ A_6^{(10)}(\lambda_1,\lambda_2,\lambda_3,\lambda_4)=A_1^{(10)}(\lambda_4,\lambda_3,\lambda_2,\lambda_1)+\tfrac{1}{6}, \nonumber\\
	&B_2^{(10)}(\lambda_1,\lambda_2,\lambda_3,\lambda_4)=B_1^{(10)}(\lambda_1,\lambda_3,\lambda_2,\lambda_4)+\tfrac{1}{18  \lambda_{12}\lambda_{34}}\left(2 - \lambda_{14} \lambda_{23} +  \lambda_{12}  \lambda_{34}  \right), \\
	&B_3^{(10)}(\lambda_1,\lambda_2,\lambda_3,\lambda_4)=B_1^{(10)}(\lambda_1,\lambda_4,\lambda_3,\lambda_2) -\tfrac{1}{18  \lambda_{12}\lambda_{34}} \left(2 - \lambda_{13} \lambda_{24} -  \lambda_{12}  \lambda_{34}  \right), \nonumber
\end{align}
\begin{align}
	&Q_1^{(11)}(\lambda_1,\lambda_2,\lambda_3,\lambda_4)=-\tfrac{1}{120}(\lambda_{12}-2 ) (2 + \lambda_{12}+5 (\lambda_{13}+1)\lambda_{23}), \nonumber\\
	&Q_2^{(11)}(\lambda_1,\lambda_2,\lambda_3,\lambda_4)=\tfrac{1}{180} (\lambda_{12}-2 ) (\lambda_{34}-2) (3 + \lambda_{23} -\lambda_{14} + 3 \lambda_{14}\lambda_{23}+ 2 \lambda_{13}\lambda_{24}), \nonumber\\ &A_2^{(11)}(\lambda_1,\lambda_2,\lambda_3,\lambda_4)=A_1^{(11)}(\lambda_1,\lambda_3,\lambda_2,\lambda_4)-\tfrac{1}{12\lambda_{12}\lambda_{14}\lambda_{34}}(2 - \lambda_{12}\lambda_{13}), \nonumber\\
	&A_3^{(11)}(\lambda_1,\lambda_2,\lambda_3,\lambda_4)=A_1^{(11)}(\lambda_1,\lambda_4,\lambda_3,\lambda_2)-\tfrac{1}{24 \lambda_{12}\lambda_{13}\lambda_{34}}(\lambda_{12}-1 )(2 + \lambda_{13} - \lambda_{34} - 2 \lambda_{13} \lambda_{34} ), \nonumber \\  &A_4^{(11)}(\lambda_1,\lambda_2,\lambda_3,\lambda_4)=A_1^{(11)}(\lambda_3,\lambda_2,\lambda_1,\lambda_4)-\tfrac{1}{24 \lambda_{12}\lambda_{24}\lambda_{34}} (\lambda_{12} - 1) (\lambda_{23}+ 2), \nonumber\\
	&A_5^{(11)}(\lambda_1,\lambda_2,\lambda_3,\lambda_4)=A_1^{(11)}(\lambda_4,\lambda_2,\lambda_3,\lambda_1),  \\ &A_6^{(11)}(\lambda_1,\lambda_2,\lambda_3,\lambda_4)=A_1^{(11)}(\lambda_4,\lambda_3,\lambda_2,\lambda_1)+\tfrac{1}{12 \lambda_{13}\lambda_{14}\lambda_{24}} (2 - \lambda_{24}\lambda_{34}), \nonumber\\
	&B_2^{(11)}(\lambda_1,\lambda_2,\lambda_3,\lambda_4)=B_1^{(11)}(\lambda_1,\lambda_3,\lambda_2,\lambda_4)-\tfrac{1}{18  \lambda_{12}\lambda_{14}\lambda_{34}}\left(2 - \lambda_{14} \lambda_{23} +  \lambda_{12}  \lambda_{34}  \right), \nonumber\\
	&B_3^{(11)}(\lambda_1,\lambda_2,\lambda_3,\lambda_4)=B_1^{(11)}(\lambda_1,\lambda_4,\lambda_3,\lambda_2) +\tfrac{1}{36  \lambda_{12}\lambda_{34}} \left(2 - \lambda_{13} \lambda_{24} -  \lambda_{12}  \lambda_{34}  \right), \nonumber
\end{align}
\begin{align}
	&Q_1^{(12)}(\lambda_1,\lambda_2,\lambda_3,\lambda_4)=-\tfrac{1}{120}(\lambda_{12}-2) (8 - \lambda_{12}+5 (\lambda_{13}-1) \lambda_{23}), \nonumber\\
	&Q_2^{(12)}(\lambda_1,\lambda_2,\lambda_3,\lambda_4)=-\tfrac{1}{180} (\lambda_{12}-2) (\lambda_{34}+2) (7 +\lambda_{12} - \lambda_{34} + 3 \lambda_{14}\lambda_{23}+ 2 \lambda_{13}\lambda_{24}), \nonumber\\ &A_2^{(12)}(\lambda_1,\lambda_2,\lambda_3,\lambda_4)=A_1^{(12)}(\lambda_1,\lambda_3,\lambda_2,\lambda_4),  ~ B_2^{(12)}(\lambda_1,\lambda_2,\lambda_3,\lambda_4)=B_1^{(12)}(\lambda_1,\lambda_3,\lambda_2,\lambda_4), \nonumber\\
	&A_3^{(12)}(\lambda_1,\lambda_2,\lambda_3,\lambda_4)=A_1^{(12)}(\lambda_1,\lambda_4,\lambda_3,\lambda_2)-\tfrac{1}{24 \lambda_{12}\lambda_{13}\lambda_{34}}(\lambda_{12}-1)(2 \!-\! \lambda_{13} \!+\! \lambda_{34} - 2 \lambda_{13} \lambda_{34} ), \nonumber \\  &A_4^{(12)}(\lambda_1,\lambda_2,\lambda_3,\lambda_4)=A_1^{(12)}(\lambda_3,\lambda_2,\lambda_1,\lambda_4)-\tfrac{1}{24 \lambda_{12}\lambda_{24}\lambda_{34}} (\lambda_{12} + 1) (\lambda_{23}+ 2), \\
	&A_5^{(12)}(\lambda_1,\lambda_2,\lambda_3,\lambda_4)=A_1^{(12)}(\lambda_4,\lambda_2,\lambda_3,\lambda_1), ~ A_6^{(12)}(\lambda_1,\lambda_2,\lambda_3,\lambda_4)=A_1^{(12)}(\lambda_4,\lambda_3,\lambda_2,\lambda_1), \nonumber\\
	&B_3^{(12)}(\lambda_1,\lambda_2,\lambda_3,\lambda_4)=B_1^{(12)}(\lambda_1,\lambda_4,\lambda_3,\lambda_2) -\tfrac{1}{36  \lambda_{12}\lambda_{13}\lambda_{34}}(\lambda_{13}-2) \left(2 - \lambda_{13} \lambda_{24} -  \lambda_{12}  \lambda_{34}  \right), \nonumber
\end{align}
\begin{align}
	&Q_1^{(13)}(\lambda_1,\lambda_2,\lambda_3,\lambda_4)=\tfrac{1}{120}(\lambda_{12}+2) (8 + \lambda_{12}+5 (\lambda_{13}+1) \lambda_{23}), \nonumber\\
	&Q_2^{(13)}(\lambda_1,\lambda_2,\lambda_3,\lambda_4)=-\tfrac{1}{180} (\lambda_{12}+2) (\lambda_{34}-2) (7  - \lambda_{12} + \lambda_{34} + 3 \lambda_{14}\lambda_{23}+ 2 \lambda_{13}\lambda_{24}), \nonumber\\ &A_2^{(13)}(\lambda_1,\lambda_2,\lambda_3,\lambda_4)=A_1^{(13)}(\lambda_1,\lambda_3,\lambda_2,\lambda_4),  ~ B_2^{(13)}(\lambda_1,\lambda_2,\lambda_3,\lambda_4)=B_1^{(13)}(\lambda_1,\lambda_3,\lambda_2,\lambda_4), \nonumber\\
	&A_3^{(13)}(\lambda_1,\lambda_2,\lambda_3,\lambda_4)=A_1^{(13)}(\lambda_1,\lambda_4,\lambda_3,\lambda_2)\!-\!\tfrac{1}{24 \lambda_{12}\lambda_{13}\lambda_{34}}(\lambda_{12}+1)(2 \!+\! \lambda_{13} \!-\! \lambda_{34} - 2 \lambda_{13} \lambda_{34} ), \nonumber \\  &A_4^{(13)}(\lambda_1,\lambda_2,\lambda_3,\lambda_4)=A_1^{(13)}(\lambda_3,\lambda_2,\lambda_1,\lambda_4)+\tfrac{1}{24 \lambda_{12}\lambda_{24}\lambda_{34}} (\lambda_{12} - 1) (\lambda_{23}- 2), \\
	&A_5^{(13)}(\lambda_1,\lambda_2,\lambda_3,\lambda_4)=A_1^{(13)}(\lambda_4,\lambda_2,\lambda_3,\lambda_1), ~ A_6^{(13)}(\lambda_1,\lambda_2,\lambda_3,\lambda_4)=A_1^{(13)}(\lambda_4,\lambda_3,\lambda_2,\lambda_1), \nonumber\\
	&B_3^{(13)}(\lambda_1,\lambda_2,\lambda_3,\lambda_4)=B_1^{(13)}(\lambda_1,\lambda_4,\lambda_3,\lambda_2) -\tfrac{1}{36  \lambda_{12}\lambda_{13}\lambda_{34}}(\lambda_{13}+2) \left(2 - \lambda_{13} \lambda_{24} -  \lambda_{12}  \lambda_{34}  \right), \nonumber
\end{align}
\begin{align}
	&Q_1^{(14)}(\lambda_1,\lambda_2,\lambda_3,\lambda_4)=\tfrac{1}{120}(\lambda_{12}+2 ) (2 - \lambda_{12}+5 (\lambda_{13}-1) \lambda_{23}), \nonumber\\
	&Q_2^{(14)}(\lambda_1,\lambda_2,\lambda_3,\lambda_4)=\tfrac{1}{180} (\lambda_{12}+2 ) (\lambda_{34}+2) (3 - \lambda_{23} +\lambda_{14} + 3 \lambda_{14}\lambda_{23}+ 2 \lambda_{13}\lambda_{24}), \nonumber\\ &A_2^{(14)}(\lambda_1,\lambda_2,\lambda_3,\lambda_4)=A_1^{(14)}(\lambda_1,\lambda_3,\lambda_2,\lambda_4)+\tfrac{1}{12\lambda_{12}\lambda_{14}\lambda_{34}}(2 - \lambda_{12}\lambda_{13}), \nonumber\\
	&A_3^{(14)}(\lambda_1,\lambda_2,\lambda_3,\lambda_4)=A_1^{(14)}(\lambda_1,\lambda_4,\lambda_3,\lambda_2)\!-\!\tfrac{1}{24 \lambda_{12}\lambda_{13}\lambda_{34}}(\lambda_{12}+1 )(2 \!-\! \lambda_{13} \!+\! \lambda_{34} \!-\! 2 \lambda_{13} \lambda_{34} ), \nonumber \\  &A_4^{(14)}(\lambda_1,\lambda_2,\lambda_3,\lambda_4)=A_1^{(14)}(\lambda_3,\lambda_2,\lambda_1,\lambda_4)+\tfrac{1}{24 \lambda_{12}\lambda_{24}\lambda_{34}} (\lambda_{12} + 1) (\lambda_{23}- 2), \nonumber\\
	&A_5^{(14)}(\lambda_1,\lambda_2,\lambda_3,\lambda_4)=A_1^{(14)}(\lambda_4,\lambda_2,\lambda_3,\lambda_1),  \\ &A_6^{(14)}(\lambda_1,\lambda_2,\lambda_3,\lambda_4)=A_1^{(14)}(\lambda_4,\lambda_3,\lambda_2,\lambda_1)-\tfrac{1}{12 \lambda_{13}\lambda_{14}\lambda_{24}} (2 - \lambda_{24}\lambda_{34}), \nonumber\\
	&B_2^{(14)}(\lambda_1,\lambda_2,\lambda_3,\lambda_4)=B_1^{(14)}(\lambda_1,\lambda_3,\lambda_2,\lambda_4)+\tfrac{1}{18  \lambda_{12}\lambda_{14}\lambda_{34}}\left(2 - \lambda_{14} \lambda_{23} +  \lambda_{12}  \lambda_{34}  \right), \nonumber\\
	&B_3^{(14)}(\lambda_1,\lambda_2,\lambda_3,\lambda_4)=B_1^{(14)}(\lambda_1,\lambda_4,\lambda_3,\lambda_2) +\tfrac{1}{36  \lambda_{12}\lambda_{34}} \left(2 - \lambda_{13} \lambda_{24} -  \lambda_{12}  \lambda_{34}  \right). \nonumber
\end{align}

\end{document}